\documentclass[%
 aip,
cha,%
 amsmath,amssymb,
preprint,%
 reprint,%
]{revtex4-1}

\usepackage{graphicx}
\usepackage{dcolumn}
\usepackage{bm}

\usepackage{nameref,hyperref}
\usepackage{amssymb,amsmath,listings, amsthm, mathtools, paralist, enumitem,tikz}
\usepackage[caption=false]{subfig}

\newcommand{\R}{\ensuremath{\mathbb{R}}}

\newcommand{\Z}{\ensuremath{\mathbb{Z}}}
\newcommand{\N}{\ensuremath{\mathbb{N}}}
\newcommand{\complex}{\mathrm{i}}

\theoremstyle{plain}
\newtheorem{thm}{Theorem}[section]

\theoremstyle{definition}
\newtheorem{defn}{Definition}[section]
\newtheorem{rem}{Remark}

\newtheorem{ex}{Example}[section]

\begin{document}

\preprint{AIP/123-QED}



\title
{Persistent homology of time-dependent functional networks constructed from coupled time series }

\author{Bernadette J. Stolz}
\author{Heather A. Harrington}%
\affiliation{ 
Mathematical Institute, University of Oxford, Oxford OX2 6GG, UK
}%

\author{Mason A. Porter}
\affiliation{Department of Mathematics, University of California Los Angeles, Los Angeles, USA}
 \affiliation{ 
Mathematical Institute, University of Oxford, Oxford OX2 6GG, UK
}%
\affiliation{%
CABDyN Complexity Centre, University of Oxford, Oxford OX1 1HP, UK
}%

\date{\today}


\begin{abstract}

We use topological data analysis to study ``functional networks'' that we construct from time-series data from both experimental and synthetic sources. We use persistent homology with a weight rank clique filtration to gain insights into these functional networks, and we use persistence landscapes to interpret our results. Our first example uses time-series output from networks of coupled Kuramoto oscillators. Our second example consists of biological data in the form of functional magnetic resonance imaging (fMRI) data that was acquired from human subjects during a simple motor-learning task in which subjects were monitored on three days in a five-day period. With these examples, we demonstrate that (1) using persistent homology to study functional networks provides fascinating insights into their properties and (2) the position of the features in a filtration can sometimes play a more vital role than persistence in the interpretation of topological features, even though conventionally the latter is used to distinguish between signal and noise.
We find that persistent homology can detect differences in synchronization patterns in our data sets over time, giving insight both on changes in community structure in the networks and on increased synchronization between brain regions that form loops in a functional network during motor learning.  For the motor-learning data, persistence landscapes also reveal that on average the majority of changes in the network loops take place on the second of the three days of the learning process.

\end{abstract}

\keywords{Persistent homology, networks, time series, fMRI, persistence landscapes, functional networks, functional brain networks, nonlinear oscillators, Kuramoto model, dynamical systems on networks}

\maketitle


\begin{quotation}

\emph{Computational topology} is a family of methods that are based on topological ideas (e.g., they often arise from algebraic topology) and give insights into topological invariants, such as connectedness or holes in high-dimensional data sets \cite{Edelsbrunner2002, Edelsbrunner2008, Edelsbrunner2010}. Such efforts have come to be called \emph{topological data analysis}, and a method known as \emph{persistent homology} (PH) has been particularly helpful for understanding shapes and their persistence over multiple scales \cite{Ghrist2014,otter2015}. Traditionally, PH has been applied to point-cloud data, though it has also been applied to networks in many applications, ranging from granular materials (see, e.g., [\onlinecite{kramar2013}]) to functional brain networks \cite{Curto2016,Giusti2016}. We employ these topological tools, which are designed to yield global, ``higher-order'' insights that go beyond revelations from pairwise connections (which are the norm in network science), in a study of functional networks constructed from both empirical and synthetic time-series data. We use persistence landscapes to show that the topological tools can (1) capture dynamics of networks constructed from the data and (2) identify mesoscale features that we relate to community structure in the associated functional networks. To help readers optimally understand these insights, we also present an intuitive introduction to PH and how to apply it to networks. 
\end{quotation}



\section{Introduction} \label{sec:Intro}

The human brain consists of approximately 100 billion neurons, whose major task is to receive, conduct, and transmit signals. Analysis of neuronal networks is crucial for understanding the human brain \cite{Bullmore2009, Bullmore2011, Sporns2014, Papo2014, Papo2014II, Betzel2016}. Every neuron consists of a cell body and one long axon, which is responsible for propagating signals to other cells \cite{Alberts2014}. Neurons or (on a larger scale) different brain regions can be construed as nodes of a network, whose edges represent either structural or functional connections between those nodes. Examining neuronal data using a network-based approach allows one to use mathematical tools from subjects such as graph theory to better understand structural and functional aspects of neuronal interactions, identify key regions in the brain that are involved in physiological and pathological processes, and compare the structure of neuronal interactions to those of other complex systems. For example, data analysis using network theory has led to the insight that the brain has underlying modular structures, with small subunits that are able to carry out specific functions while minimally influencing other parts of the brain \cite{Bullmore2009, Bullmore2011II, Papo2014}. 

The standard methods from network theory are based on pairwise connections, which one can use to study microscale, mesoscale, and macroscale structures~\cite{Newman2013}. An alternative approach for studying networks\footnote{In the present paper, we use the terms ``network'' and ``graphs'' synonymously, although the former is often used in a way that includes structures that are more complicated than ordinary graphs.} is to use methods from \emph{computational topology}, which explicitly incorporates ``higher-order'' structures beyond pairwise connections and includes algorithmic methods for understanding topological invariants such as connectedness, loops, or holes in high-dimensional data structures \cite{Edelsbrunner2002, Edelsbrunner2008, Edelsbrunner2010}(see Section \ref{sec:Homology and Betti numbers}). Although one can also represent higher-order structures using formalisms such as hypergraphs~\cite{Bollobas1998} (see, e.g., a recent paper\cite{Bassett2014II} by Bassett et al.), those other approaches may not be the most convenient means for optimally conveying information about the shape or scale of mesoscale structures in a network. Other recent work concerns clustering in networks using higher-order structures \cite{benson2016}.

Methods from computational topology enable one to understand global low-dimensional structures in networks, and they have led to insights in an increasingly large number of applications \cite{otter2015} in diverse topics, ranging from granular materials \cite{kramar2013} and contagions on networks \cite{Taylor2015} to path planning\cite{bgk2015} and collective behavior in animals \cite{topaz2014}. In particular, \emph{persistent homology} (PH), a mathematical formalism to explore the persistence of topological structures in data sets, has become increasingly prominent in neuroscience in the last few years \cite{Curto2016,Giusti2016}. Among other applications, it has been used to determine differences in brain networks of children with hyperactivity disorders and autism spectrum in comparison to normal situations \cite{Lee2011}, study the effect of the psychoactive component of ``magic mushrooms'' (psilocybin mushrooms) on functional brain networks of humans \cite{Petri2014}, analyze covariates that influence neural spike-train data \cite{Spreemann2015}, and study structural and functional organization of neural microcircuits \cite{Dlotko2016}. Other neuronal applications have included consideration of place cells in the hippocampus of rats during spatial navigation \cite{Curto2008, Dabaghian2012, Giusti2015}, analysis of mathematical models of transient hippocampal networks\cite{Dabaghian2016}, and a demonstration that topological features of networks of brain arteries in humans are correlated with their age \cite{Bendich2014}. We also note that PH is not the only topological method that has been used to study the human brain or time series. More than fifty years ago, for example, Zeeman~\cite{Zeeman1962} used tolerance spaces and Vietoris homology theory to study aspects of visual perception. In the 1990s, Muldoon et al.~\cite{Muldoon1993} developed a method to study the topology of manifolds that underlie time-series data.

In the present investigation, we use time-series data to construct so-called \textit{functional networks}~\cite{Bullmore2011,Bullmore2009,Sporns2015,Petersen2015} (but note that one can study coupled time series using a variety of different approaches~\cite{Freitas2015,Tirabassi2014,sun2014,small2016}). Functional brain networks consist of a set of nodes (e.g., brain regions) and a set of weighted edges between nodes, where the edge weights quantify the similarity of the associated time series according to a chosen measure. A functional network contrasts with a ``structural network,'' which refers to underlying physical connections (e.g., anatomical connections) between nodes. For example, neurons are connected to each other in structural networks, but one can analyze the similarity in their firing patterns through functional networks. We use the term ``functional network'' in a more general way: by constructing a matrix of similarities between coupled time series using some measure (and enforcing the diagonal entries to be $0$), one obtains a functional network whose weighted adjacency matrix (sometimes also called an ``association matrix'') $\tilde A = (\tilde a_{ij})_{i,j = 1}^N$ has elements that indicate the similarity between the time series of entities $i$ and $j$. Studying functional networks is common in neuroscience, and they are also used in a wealth of other applications (e.g., finance \cite{fenn2009}, voting among legislators \cite{waugh2009}, and climate \cite{donges2009}). Importantly, the times series can come either from empirical data or from the output of a dynamical system (or stochastic process), and the latter is helpful for validating methods for network analysis \cite{Bassett2013}. In our paper, we will consider times series either from coupled oscillators (i.e., as the output of a dynamical system) or from a set of spatially distinct brain regions defined by a fixed anatomical atlas.  In the context of functional brain networks, the adjacency-matrix element $a_{ij}$ arises as a measure of ``functional connectivity'' (i.e., behavioral similarity) between the time series for nodes (i.e., brain regions) $i$ and $j$. There are many different ways to measure similarity of times series \cite{Smith2011,Zhou2009,Bullmore2011}, and that can be a major issue when it comes to interpreting results. Comparing the networks that arise from different similarity measures is beyond the scope of our work, so we will simply use two common measures (pairwise synchrony and wavelet coherence) of time-series similarity. However, the methods that we employ can be applied to functional networks that are constructed using any measure of similarity between time series.

In many studies based on experimental data, functional networks are used to construct binary graphs (i.e., unweighted graphs) \cite{Bullmore2011}. To do this, one typically applies a global threshold $\xi \in \R^+$ to a weighted adjacency matrix to obtain a binary adjacency matrix $A = (a_{ij})_{i,j = 1}^N$ associated with an unweighted graph. The adjacency-matrix elements are then
\begin{equation}
	a_{ij} = \begin{dcases}
		1\,, \ \ \text{if  } \tilde a_{ij} \geq \xi\,, \\
		0\,, \ \ \text{otherwise\,.}
	\end{dcases}
\end{equation}
The choice of threshold has a strong influence on the resulting matrix, and it thereby exerts a major influence on the structure of the associated graph \cite{Bullmore2011}. Some approaches to address this issue include determining a single ``optimal'' threshold, thresholding the weighted adjacency matrix at different values\cite{Serrano2009,Alexander-Bloch2010}, examining the network properties as a function of threshold, or not thresholding at all and considering the weighted adjacency matrix itself \cite{Bullmore2011,Papo2014}. (One can also threshold a weighted adjacency matrix by setting sufficiently small entries to $0$ but keeping the values of the other entries.) If one is thresholding and binarizing data, there is no guarantee that there exists an interval of thresholds that yield networks with qualitatively similar properties, and arbitrarily throwing away data can be problematic even when such intervals do exist. For example, parameters such as graph size (i.e., number of nodes) need to be taken into account when interpreting results on thresholded, binarized networks \cite{Fallani2014}. An advantage of using persistent homology is that one can examine a graph ``filtration'' (see Section \ref{sec:Filtrations}) generated by multiple --- ideally all --- possible global thresholds and systematically analyze the persistence of topological features across these thresholds. Such a filtration can also be created using decreasing local thresholds.
  
In our topological analysis, we focus on ``loops'' in a network. A loop in a graph is a set of at least four edges that are connected in a way that forms a topological circle. Loops are thus 1-dimensional topological features. We choose to focus on loops rather than features with dimension $0$, which correspond to connected components of a graph, are topologically simpler, and can be studied using many other approaches (e.g., through the number of $0$ elements in the spectrum of the combinatorial graph Laplacian~\cite{Bollobas1998}). It has been demonstrated in other applications (e.g., contagions on networks~\cite{Taylor2015}) that loops are important topological features of graphs, and a recent study~\cite{Sizemore2016} demonstrated the importance of loops (and related higher-dimensional objects) in structural neuronal networks. Structural and functional neuronal networks are related and share some network features~\cite{Bullmore2009}, so we expect loops to provide interesting insights.

The remainder of our paper is organized as follows. In Section~\ref{sec:CompTop}, we give a brief and intuitive introduction to persistent homology, the weight rank clique filtration, and persistence landscapes. In Section~\ref{sec:Ex1}, we introduce our first example, the Kuramoto model of nonlinearly coupled oscillators; and we present results from our application of persistent homology to time-series data produced by coupled Kuramoto oscillators. In Section~\ref{sec:Ex2}, we introduce and analyze our second example, which consists of time-series functional magnetic resonance imaging (fMRI) data from experiments of humans performing a simple motor task. We present our conclusions in Section~\ref{sec:ExConcDisc}, and we provide a mathematical introduction to persistent homology in the Supplementary Information.


\section{Persistent homology} \label{sec:CompTop}

Persistent homology (PH)\cite{Edelsbrunner2002, Edelsbrunner2008, Edelsbrunner2010} is a method from computational topology that quantifies global topological structures (e.g., connectedness and holes) in high-dimensional data. One can think of PH as looking at the ``shape'' of data in a given dimension using a set of different lenses. Each lens conveys topological features inside data at a different resolution. One then construes structures that persist over a range of different lenses to represent a significant feature of the data. Structures that are observed only through a small number of lenses are commonly construed as noise \cite{Ghrist2008,Carlsson2009}, especially in settings where the data are sampled from a manifold. For empirical data, the relationship between small persistence of a feature and whether it constitutes noise in a data set rather than signal has not yet been verified statistically, but we will illustrate a situation in which some short-lived structures represent important features and possibly genuine geometrical (not just topological) features of data in Sections~\ref{sec:Ex1} and ~\ref{sec:Ex2}.

In this section, we provide an intuitive introduction to the mathematical concepts behind PH. In Supplementary Information, we give a mathematically rigorous introduction (including precise definitions).


\subsection{\label{sec:Simplicial complexes}Simplicial complexes}

One can study the properties of a topological space\cite{Kosniowski1980, Munkres2000} by partitioning it into smaller and topologically simpler pieces, which when reassembled include the same aggregate topological information as the original space.  The most trivial topological space $X = \{ \emptyset, x \}$ consists of the empty set $\emptyset$ and a single point $x$. If we want to simplify the description of the topological properties of $X$, we would simply choose a single node to represent it. However, a node or even a collection of nodes does not allow one to capture the topological properties of more complicated spaces, such as a $2$-sphere or the surface of the earth. In such cases, one needs a simple object that carries the information that the space is connected but also encloses a hole. For example, one could use a tetrahedron, which is an example of a mathematical object called a ``simplex.''

\begin{figure}[h!]
\centering\includegraphics[width=1.7cm]{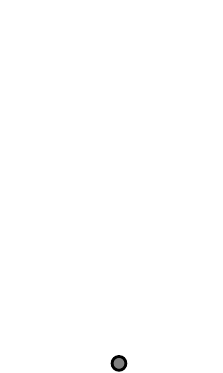} \ \ \ \
\centering\includegraphics[width=0.65cm]{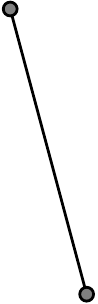} \ \ \ \ \
\centering\includegraphics[width=1.7cm]{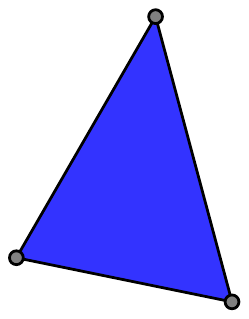} \ \ \ \
\centering\includegraphics[width=2.25cm]{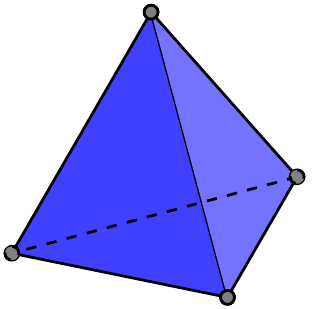} \ \ \ \
\caption[Examples of low-dimensional simplices.]{Examples of (from left to right) a 0-simplex, a 1-simplex, a 2-simplex, and a 3-simplex. [We adapted this figure from [\onlinecite{Edelsbrunner2010}].]}\label{fig:Tetrahedron}
\end{figure}

The building blocks that one uses to approximate topological spaces are called \emph{$k$-simplices}, where the parameter $k$ indicates the dimension of the simplex. Every $k$-simplex contains $k+1$ independent nodes: a point is a $0$-simplex, an edge is a $1$-simplex, a triangle is a $2$-simplex, and a tetrahedron is a $3$-simplex (see Fig.\ref{fig:Tetrahedron}). Observe that the lower-dimensional simplices are contained in the higher-dimensional simplices. This allows one to build higher-dimensional simplices using lower-dimensional ones. The lower-dimensional simplices form so-called \textit{faces} of the associated higher-dimensional objects. 

\begin{figure*}[t!]
\subfloat[]{%
  \includegraphics[clip, width = 0.11\textwidth]{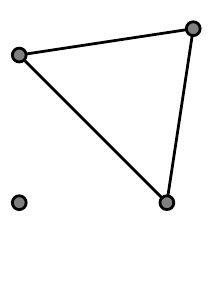}%
}
~ 
~ 
~ 
~ 
\subfloat[]{%
  \includegraphics[clip,width = 0.3\textwidth]{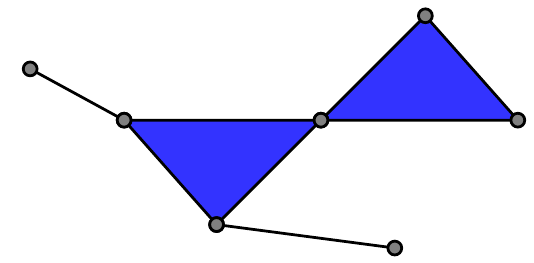}%
}
~ 
~ 
~ 
~ 
~ 
\subfloat[]{%
  \includegraphics[clip,width = 0.11\textwidth]{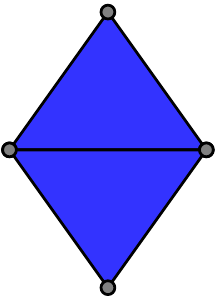}%
}
~ 
~ 
~ 
~ 
~ 
\subfloat[]{%
  \includegraphics[clip,width = 0.125\textwidth]{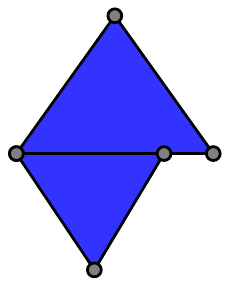}%
}
\caption{\label{fig:Simplicial}Panels (a)--(c) give examples of simplicial complexes, and panel (d) gives an example of an object that is not a simplicial complex. The blue color indicates a $2$-simplex. Example (a) illustrates that simplicial complexes are not necessarily also simplices. The three edges do not form a $2$-simplex; instead, they form a simplicial complex that consists of $1$-simplices. In examples (b) and (c), all $1$-simplices and $2$-simplices are connected by $0$-simplices. Example (d) is a collection of simplices that violates the definition of a simplicial complex, because the intersection between the two triangles does not consist of a complete edge that is shared by both simplices (as it includes only approximately 75\% of the base edge of the upper triangle). Note that any combination of the three simplicial complexes (a), (b), and (c) is also a simplicial complex.}
\end{figure*}

One combines different simplices into a \emph{simplicial complex} to capture different aspects of a topological space. For every simplex that is part of a simplicial complex, we demand that all of its faces are also contained in the simplicial complex. Additionally, two simplices that are part of a simplicial complex are allowed to intersect only in common faces. In Fig.~\ref{fig:Simplicial}, we show several examples of simplicial complexes and one example that is not a simplicial complex.

We take the \emph{dimension} of a simplicial complex to be the dimension of its highest-dimensional simplex. One can use simplicial complexes to represent topological spaces if and only if there exists a continuous deformation that can stretch and bend the simplicial complex into the topological space, and only then are topological properties of the topological space preserved by the simplicial complex.


\subsection{Homology and Betti numbers} \label{sec:Homology and Betti numbers}

\begin{figure}[h!]
 \includegraphics[width=.25\textwidth]{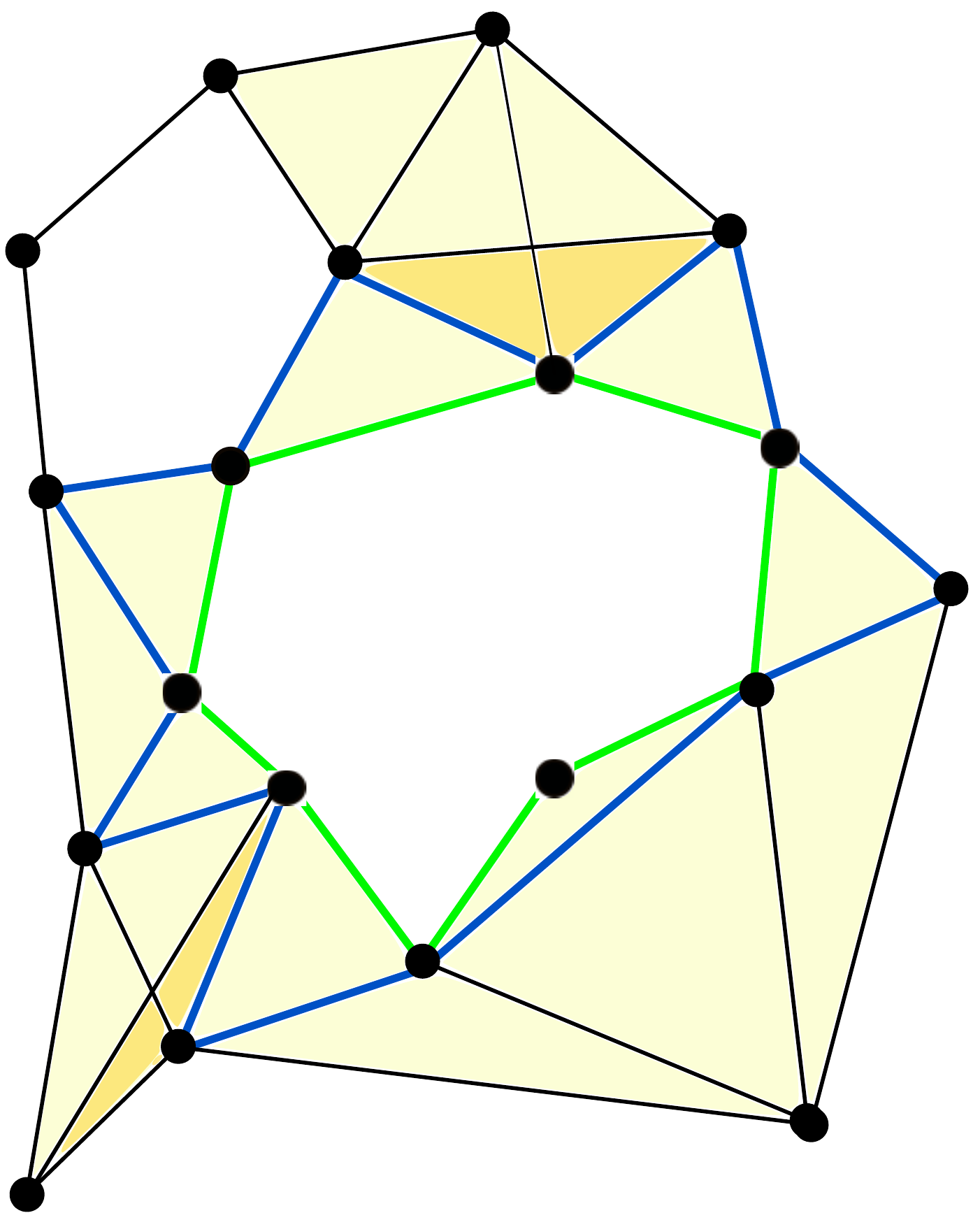}
 \caption{Example of a $1$-dimensional loop in a simplicial complex. The green and the blue loop both surround the same hole and are therefore considered to be representatives of the same homology class.} \label{Fig:Loop}
 \end{figure}

If one is interested in the nature of a simplicial complex of dimension $k$, one can either consider the full complex, which can be very large, or one can examine different subsets of simplices that are contained in the complex. For example, the set of all $1$-simplices consists of a collection of edges, some of which may be connected or even form a loop. However, one can consider a range of different topological features --- e.g., in some cases, a collection of edges surrounding a hole or void could be more interesting than individual edges --- and one typically seeks features that are invariant if one squeezes or stretches the edges. \emph{Homology} is a formal way to quantitatively detect topological invariants in a given dimension to give insight into the nature of a topological space. By using homology, one can, for example, distinguish a $2$-sphere from a torus. For a simplicial complex of dimension $k$, one can define a vector space known as the $p$th \emph{homology group} for every dimension $p \in \{0, \dots, k \}$.  In dimension $1$, for example, the elements of the homology group are called ``loops.'' The elements of the homology group can be divided into different \emph{homology classes}, which each represent a hole in the topological space. For instance, in dimension $1$, loops in the same homology class all surround the same $1$-dimensional hole. We give an example of two loops that surround the same hole in Fig.~\ref{Fig:Loop}. The homology classes yield a family of vector spaces, whose dimensions are called \emph{Betti numbers}, associated to a simplicial complex. One can interpret the first three Betti numbers, $\beta_0$, $\beta_1$, and $\beta_2$, to represent, respectively, the number of connected components, the number of $1$-dimensional holes, and the number of $2$-dimensional holes in a simplicial complex. As we pointed out in Section \ref{sec:Intro}, we focus on loops (i.e., $\beta_1$) in our network analysis rather than $\beta_0$, which corresponds to the number of connected components in a graph. One can study connected components in graphs using many other approaches, such as by calculating the number of $0$ eigenvalues in the spectrum of the combinatorial graph Laplacian~\cite{Bollobas1998}.

\subsection{Filtrations} \label{sec:Filtrations}

Although homology gives information about a single simplicial complex, it is typically more relevant to study topological features across sequences (called \emph{filtrations}) of simplicial complexes. A filtration~\cite{Carlsson2009, Edelsbrunner2008, Ghrist2008} of a simplicial complex $\Sigma$ is a sequence of embedded simplicial complexes,
\begin{equation}
	\emptyset = \Sigma_0 \subseteq \Sigma_1 \subseteq \Sigma_2 \subseteq \dots \subseteq \Sigma_k = \Sigma\,,
\end{equation}
starting with the empty complex and ending with the entire simplicial complex. One can use homology to study topological features (e.g., $1$-loops) in every step of the filtration and determine how persistent they are with respect to a given filtration. A topological feature $h$ is \emph{born} at $\Sigma_m$ if $\Sigma_m$ is the first simplicial complex in the filtration to contain the feature. Similarly, a topological feature \emph{dies} in $\Sigma_n$ if it is present in $\Sigma_{n-1}$ but not in $\Sigma_n$. One then defines the \emph{persistence} $p$ of the topological feature as 
\begin{equation*}
	p = n-m\,. 
\end{equation*}	
Persistence was first used as a measure to rank topological features by their lifetime\cite{Edelsbrunner2002} in a filtration in $\R^3$.

There are many ways to define simplicial complexes and filtrations on weighted graphs, and the choice of filtration tends to be motivated either by the type of questions to be answered or by the consideration of computation time.


\subsubsection{Weight rank clique filtration} \label{sec:WRCF}

Although we focus on network data, we note that PH has been applied much more often to data in the form of point clouds \cite{Carlsson2009, Ghrist2008}. The simplest way to create a sequence of embedded graphs (e.g., a filtration) from a weighted network is to filter by weights \cite{Lee2012}. To do this, one creates a sequence of embedded (binary) graphs by ranking all edge weights $\nu_t$ in descending order. In filtration step $t$, one retains an edge if and only if its weight is at least $\nu_t$. To construct the filtration, one repeats this procedure until the graph is complete in the last step. Using this method, only $0$-simplices (i.e., nodes) and $1$-simplices (i.e., edges) are present in the filtration. The weight rank clique filtration (WRCF) ~\cite{Petri2013}, which we will use in our analysis and which has been applied previously for examining weighted neuronal networks \cite{Petri2013, Petri2014, Giusti2015}, extends this definition to include higher-dimensional simplices. One constructs a WRCF as follows:
\begin{enumerate}
\item Define filtration step $0$ as the set of all nodes.
\item Rank all edge weights $\{ \nu_{1},\dots, \nu_{\text{end}} \}$, with $\nu_{1} = \nu_{\mathrm{max}}$ and $\nu_{\text{end}} = \nu_{\mathrm{min}}$. (We will use $\tau$ to denote the number of distinct weights in a graph.)
\item In filtration step $t$, threshold the graph at weight $\nu_t$ 
to create a binary graph.
\item Find all maximal $c$-cliques for $c \in \N$, and define them to be $c$-simplices.
\end{enumerate}
This is a valid simplicial complex: every $(c+1)$-clique in the graph guarantees the existence of a $c$-face on that clique, because cliques are closed under both intersection and taking subsets. Consequently, they satisfy the requirements for a simplicial complex. This type of simplicial complex on a graph is called a \textit{clique complex}. 

\begin{figure}[h!]
\centering\includegraphics[width=0.47\textwidth]{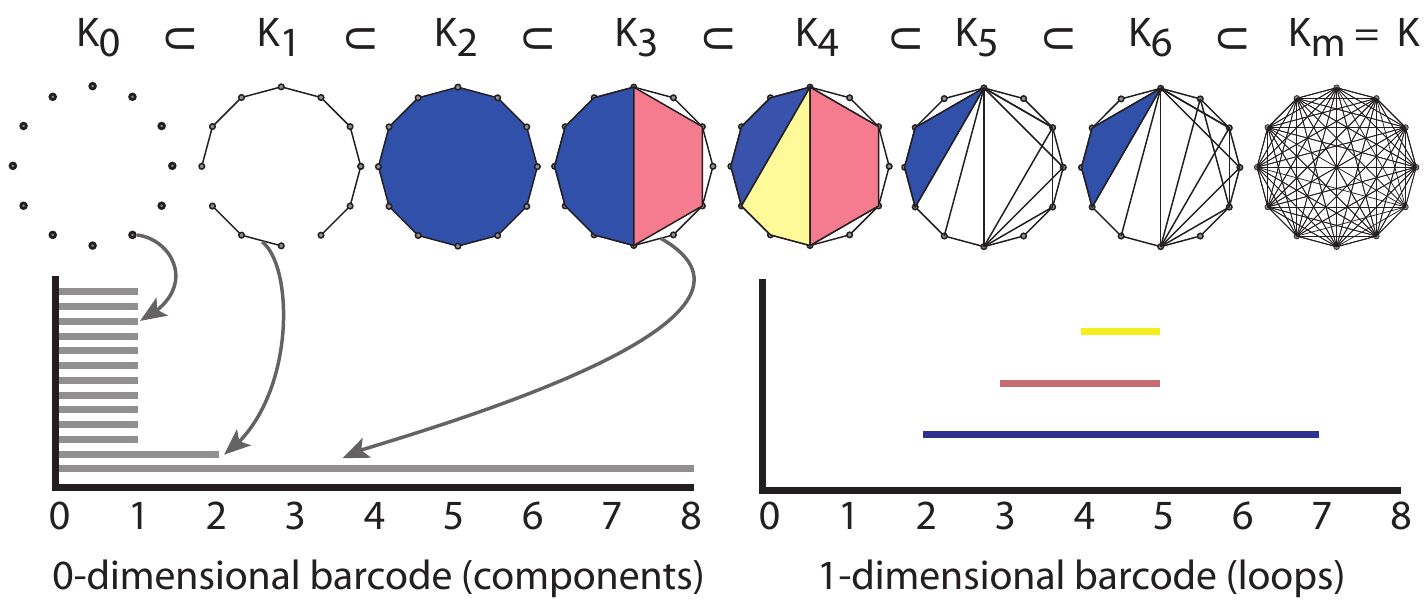}
\caption{Example of a weight rank clique filtration (WRCF) and the corresponding $0$-dimensional and $1$-dimensional barcodes. The barcode of dimension $0$ indicates the connected components in every filtration step. When two components merge into one connected component, one of the bars that represents the original components dies in the barcode; the other continues to the next filtration step and now represents the newly-formed component. In filtration step $0$, every node is a separate component, resulting in 12 bars in the barcode. The nodes are joined to become two components in filtration step 1, and they then become a single component in step 2. In dimension $1$, we observe that as more edges are added to the filtration, the loop surrounding the blue hole born in filtration step $2$ is divided first into two holes and subsequently into three holes before it is completely covered by $2$-simplices and dies in filtration step $7$. The colors of the bars indicate which loop they represent.
}\label{Figure WRCF Example} 
\end{figure}

One can visualize the persistence of homology classes of a filtration of a simplicial complex using \emph{barcodes}~\cite{Ghrist2008}. 
A barcode for a given dimension is a collection $\{b_l,d_l \}_{i = 1}^m$ of intervals, where every interval $(b_l,d_l)$ represents a topological feature $l$ of the given dimension (examples of such features include connected components and loops), $b_l$ denotes the birth time of feature $l$ with respect to the filtration step, and $d_l$ denotes its death time. The length $d_l - b_l$ of the bar measures the persistence of the feature. In Fig.~\ref{Figure WRCF Example}, we show an example of a WRCF and its corresponding barcode.

\subsection{Persistence landscapes} \label{sec:PersistenceLandscapes}

As an alternative topological summary to barcodes, one can use \emph{persistence landscapes}~\cite{Bubenik2015I,Bubenik2015}, which consist of piecewise-linear functions in a separable Banach space. For a given barcode interval $(b,d)$, one defines the function
\begin{equation}
f_{(b,d)} = \begin{cases} 0 \,, & \text{if } x \notin (b,d) \,, \\
		x - b 	\,,	&\text{if } x \in (b,\frac{b+d}{2}] \,, \\
		-x + d \,,	&\text{if } x \in (\frac{b+d}{2},d ) \,.
		\end{cases}
\end{equation}
For a barcode $\{b_l,d_l \}_{i = 1}^m$ and $q \geq 0$, the \emph{$q$}th \emph{persistence landscape} is given by the set of functions 
\begin{align}
	& \lambda_q: \R \rightarrow \R\,, \\
	& \lambda_q(x) =  \text{\emph{$q$}th-largest value of } \{f_{(b_l,d_l)}(x) \}_{l = 1}^m \,. \nonumber
\end{align}
If the \emph{$q$}th-largest value does not exist, then $\lambda_q(x) = 0$. One can think of the $0$th persistence landscape as being the outline of the collection of peaks created by the images of the collection of functions $f$ associated to a barcode. To obtain the $1$st persistence landscape, one peels away this topmost ``layer'' of peaks and then considers the outline of the remaining collection of peaks. This gives the $1$st persistence landscape, and one continues in this manner to obtain subsequent persistence landscapes. The \emph{persistence landscape} $\lambda$ of the barcode $\{b_l,d_l \}_{l = 1}^m$ is then defined as the sequence $\{ \lambda_q \}$ of functions $\lambda_q$. 

\begin{figure*}[ht!]
\centering\includegraphics[width=\textwidth]{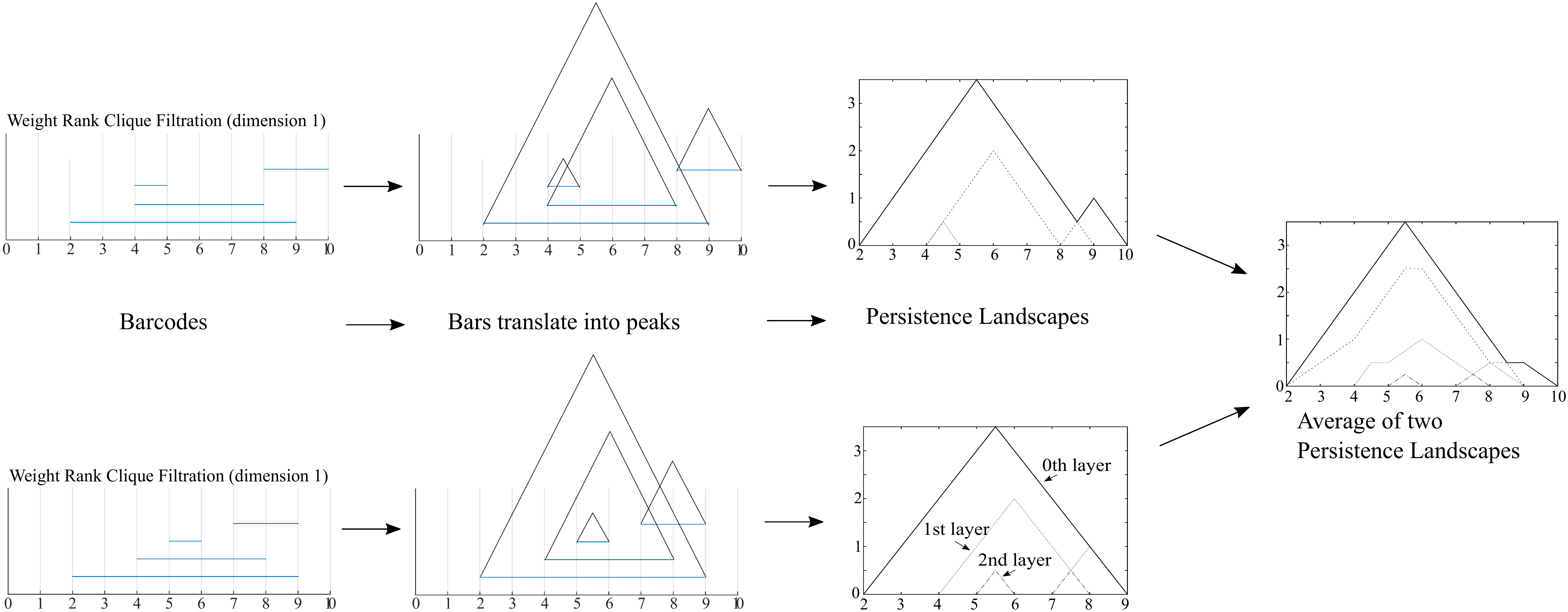}
\caption{Visualization of the relationship between barcodes and an average persistence landscape. To obtain a landscape from a barcode, one replaces every bar of the barcode by a peak, whose height is proportional the persistence of the bar. In the landscape, we translate all peaks so that they touch the horizontal axis. The persistence landscape consists of different layers, where the $q$th layer corresponds to the $q$th-largest function value across the collection of peak functions. One creates an average of two landscapes by taking the mean over the function values in every layer.
}\label{Fig: Landscapes} 
\end{figure*}

Even though persistence landscapes visualize the same information as barcodes and one can construct a 
bijective correspondence between the two objects, the former have distinct advantages over the latter. 
For example, one can calculate a unique ``average landscape'' for a set of persistence landscapes by taking the mean over the function values for every landscape layer.
This is not possible for barcodes, as they are not elements of a Banach space. For an average landscape, it is thus not possible to find a corresponding average barcode. We show a schematic illustration on how to obtain an average persistence landscape in Fig.~\ref{Fig: Landscapes}.

One can also define $L^p$ distances between two (average) landscapes and thereby use a variety of statistical tools~\cite{Bubenik2015I}. This allows one to compare multiple groups of barcodes by calculating a measure of pairwise similarity between them. Persistence landscapes have been used to study conformational changes in protein binding sites\cite{Kovacev2015}, phase separation in binary metal alloys~\cite{Dlotko2016II} and music audio signals~\cite{Liu2016}.


\subsection{Computational tools} \label{sec:ComputationalTools}

For our PH calculations, we use {\sc Matlab} code that we construct using {\sc javaPlex} \cite{javaPlex, Adams2014}, a software package for persistent homology. For the WRCFs, we also use a maximal clique-finding algorithm from the Mathworks library \cite{Wildmann2011} based on the Bron--Kerbosch algorithm, which is the most efficient algorithm known for this problem. For statistical analysis and interpretation of our barcodes, we apply the {\sc Persistence landscapes toolbox}~\cite{Bubenik2015}.


\section{Example I: Coupled Kuramoto oscillators} \label{sec:Ex1}

\subsection{The Kuramoto model}

The Kuramoto model~\cite{Kuramoto1984,Strogatz2000,Arenas2008,Rodrigues2016,Gupta2014} is a well-studied model for a set of coupled phase oscillators with distinct natural frequencies that are drawn from a prescribed distribution. The model was developed in the 1970s to understand collective synchronization in a large system of oscillators. It has subsequently been used as a toy model by many neuroscientists (as well as scholars in many other areas), as some of the characteristics of its synchronization patterns resemble some of the ones in neuronal communities \cite{Velazquez2006, Breakspear2010, Ashwin2016, Lee2016}. The Kuramoto model and its generalizations have also been applied to numerous other applications in chemistry, biology, and other disciplines \cite{Arenas2008,Rodrigues2016,porter2016}. 

When all oscillators are coupled to each other, the Kuramoto model is most commonly written as~\cite{Strogatz2000,Rodrigues2016}
\begin{equation}
	\frac{d\theta_i}{dt} = \omega_i + \frac{K}{N} \sum^N_{j = 1} \sin (\theta_j - \theta_i)\,, \ \ \ i \in \{1,\dots,N\}\,,\label{eq:Kuramoto}
\end{equation}
where $\theta_i$ denotes the phase of oscillator $i$, the parameter $\omega_i$ is its natural frequency, $K \geq 0$ parametrizes the coupling strength between different oscillators, and $N$ is the number of oscillators in the model. The normalization factor $\frac{1}{N}$ ensures that the equations are bounded as $N \rightarrow \infty$. The distribution from which the frequencies $\omega_i$ are drawn is usually assumed to be unimodal and symmetric about its mean frequency, which can be set to $0$ due to the rotational symmetry of the model (because Eq.~(\ref{eq:Kuramoto}) is invariant under translation of $\theta_i$). The parameter $\omega_i$ then denotes the deviation from the mean frequency.

We also adapt Eq.~(\ref{eq:Kuramoto}) to create a network of $N$ oscillators with uniform coupling between the oscillators~\cite{Arenas2006, Arenas2008,Bassett2013,Tirabassi2014,Rodrigues2016}. We consider the following generalized version of Eq.~(\ref{eq:Kuramoto}):
\begin{equation}
	\frac{d\theta_i}{dt} = \omega_i + \sum^N_{j = 1} \kappa A_{ij} \sin (\theta_j - \theta_i)\,, \ \ \  i \in \{1,\dots,N\}\,, \label{eq:KuramotoNet}
\end{equation}
where $\kappa \geq 0$ denotes the normalized coupling strength and the entries of the coupling matrix $A = (A_{ij})_{i,j = 1}^N$ indicate whether oscillators $i$ and $j$ are coupled. That is, $A$ is an unweighted adjacency matrix, and $A_{ij} = 1$ for coupled oscillators and $A_{ij} = 0$ for uncoupled oscillators. The coupling matrix $A$ thereby imposes a ``structural network'' between the oscillators. One can further generalize Eq.~(\ref{eq:KuramotoNet}) by using heterogeneous coupling strengths $\kappa_{ij}$ or by considering functions other than sine on the right-hand side. 

We divide the oscillators into 8 separate communities \footnote{In this context, we use the term ``community'' to indicate a set of densely-connected nodes with sparse connections to other nodes outside of this set. There are also other uses of the term, and community structure is a popular subject in network science \cite{Porter2009,fortunato2016}.} of 16 distinct oscillators each, and we suppose that every oscillator has exactly $14$ connections, 13 of which are with oscillators in the same community and 1 of which is to an oscillator outside the community. As in Bassett et al.\cite{Bassett2013}, we choose a coupling strength of $\kappa = 0.2$, consider a network with $N = 128$ oscillators, and suppose that the $i$th natural frequency $\omega_i \sim \mathcal{N}(0,1)$. (That is, we draw natural frequencies from a Gaussian distribution with mean $0$ and standard deviation $1$.) However, our network architecture differs somewhat from that in Bassett et al.~\cite{Bassett2013}, where every oscillator had at least 13 connections inside its community and at least 1 connection outside its community. 

We simulate the basic Kuramoto model using the Runge--Kutta {\sc Matlab} solver {\sc ODE45} (with an integration time interval of $[0, T_{\text{max}}]$, where $T_{\text{max}}=10$)\footnote{We use an input time step of $\Delta t = 0.02$, but we note that {\sc ODE45} uses an adaptive step size.}. We observe the system for $M = 500$ time steps in total (including the initial time step) and obtain time series $\mathcal{T}_i = (\theta_i(t_0), \dots, \theta_i(t_{499}))$ as the output of the model for every oscillator $\theta_i$. 
Kuramoto oscillators with a similar imposed community structure were demonstrated previously to initially synchronize rapidly within their communities, followed by a phase of global synchronization in an entire network~\cite{Bassett2013}. (There have also been other studies of community structure via synchronization of Kuramoto oscillators\cite{Arenas2006,stout2011}.) To study the dynamics of the coupled Kuramoto oscillators, we follow the approach of Bassett et al.~\cite{Bassett2013} and partition the time series into two time regimes, which we denote by $\hat{k} = 1$ and $\hat{k} = 2$. In our example, these time regimes each consist of $250$ time steps. (Reference~\onlinecite{Bassett2013} also split their time series into two equal parts, but their time series consist of 100 time steps in total rather than 500.)

To quantify the pairwise synchrony of two oscillators $i$ and $j$, we use the local measure~\cite{Arenas2006, Bassett2013}
\begin{equation}
	\phi_{ij}^{\hat{k}} = \left\langle \left| \cos \left( \mathcal{T}_i^{\hat{k}} - \mathcal{T}_j^{\hat{k}}\right) \right| \right\rangle\,,\label{eq:simMeasure}
\end{equation}
where the angular brackets indicate that we take a mean over 20 simulations. We use the absolute value both to facilitate comparison with Arenas et al.~\cite{Arenas2006} and Bassett et al.~\cite{Bassett2013} (by making the same choice that they made) and to avoid negative values, which can complicate interpretation and pose other difficulties in network analysis \cite{fenn2009,traag2009,Smith2011}.

\begin{figure}[h!]
\centering\includegraphics[width=8.5cm]{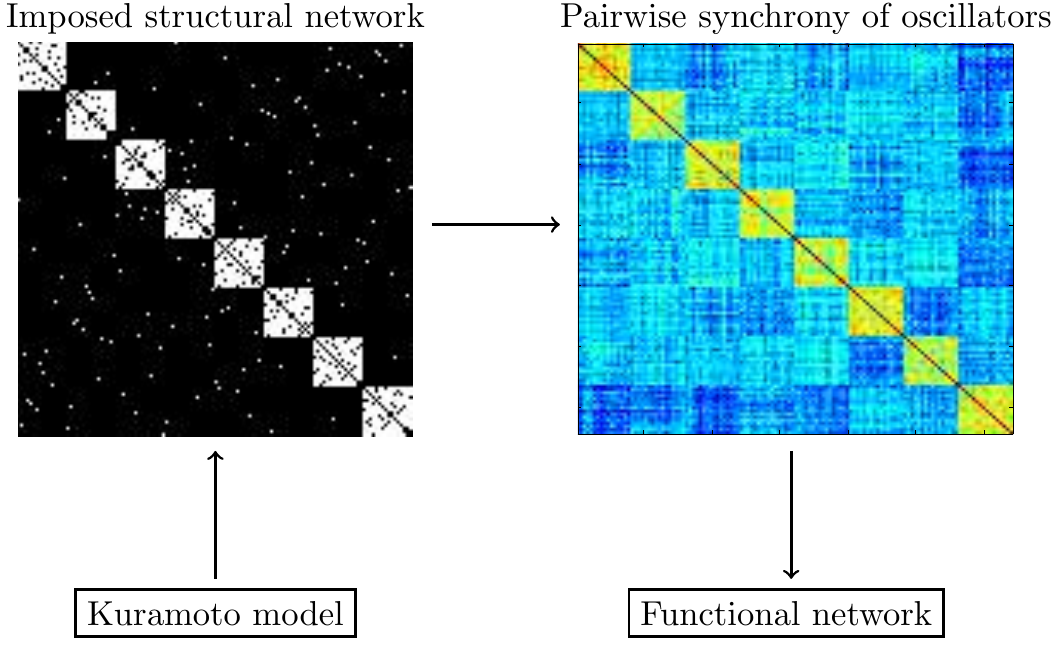}
\caption{We construct a structural network for coupled Kuramoto oscillators by grouping the oscillators into 8 separate communities. Oscillators are coupled predominantly to other oscillators in their community, and they are coupled only very sparsely to oscillators outside their community. We use the time-series output of a simulation of the Kuramoto model to create a functional network based on the similarity of the time series of individual oscillators. We use the measure of similarity in Eq.~\eqref{eq:simMeasure}.}\label{fig:KuraPipeline}
\end{figure}

In each simulation, we choose the initial values for the phases $\theta_i$ from a uniform distribution on $[0,2\pi)$ and draw the natural frequencies $\omega_i$ from $\mathcal{N}(0,1)$. We apply the same underlying coupling matrix $A = (A_{ij})_{i,j = 1}^N$ for all 20 simulations and then use the values $\phi_{ij}$ to define the edge weights in the fully connected, weighted network of Kuramoto oscillators for each time regime. We also study a network based on one full time regime that consists of $500$ time steps. In analogy to neuronal networks, we call these networks ``functional networks.'' In Fig.~\ref{fig:KuraPipeline}, we illustrate our pipeline for creating a functional network from the output of a simulation of the Kuramoto model.

\subsection{Null models for the Kuramoto data}\label{sec:KuramotoNull}

To assess whether our observations illustrate meaningful dynamics of the Kuramoto model or whether they can be explained by a random process, we consider two different null models based on the time-series output. In the first null model, which we call the ``simple null model,'' we reassign the order of the time series for every oscillator according to a uniform distribution before computing the similarity measure with Eq.~(\ref{eq:simMeasure}). The second null model, which we call the ``Fourier null model,'' is based on creating surrogate data using a discrete Fourier transformation. This approach~\cite{Prichard1994} has the advantage of preserving not only the mean and the variance of the original time series but also the linear autocorrelations and cross correlations between the different time series. 

To construct the Fourier null model, we start by taking the discrete Fourier transform 
\begin{equation}
	\hat{\mathcal{T}}_n = \frac{1}{\sqrt{\mu}}\sum_{m = 0}^{\mu-1} \mathcal{T}_m e^{\frac{2\pi \complex nm}{\mu}}
\end{equation}
of a time-series vector $\mathcal{T} = (\theta(t_0), \dots, \theta(t_{\mu}))$ of length~$\mu$. In our case, $\mu = 250$ or $\mu = 500$, depending on whether we are examining two different time regimes or just one. We then construct surrogate data by multiplying the Fourier transform $\hat{\mathcal{T}}_n$ by phases $a_n$ chosen uniformly at random from the interval $[0,2\pi)$, aside from the constraint that they must satisfy the following symmetry property: for every $n \leq \mu$, there exists $\tilde{n}$ such that $a_n = -a_{\tilde{n}}$. 
This symmetry ensures that the inverse Fourier transform yields real values. The surrogate data $\sigma = (\sigma_1,\dots,\sigma_{\mu})$ are thus given by
\begin{equation}
	\sigma_m = \frac{1}{\sqrt{\mu}}\sum_{n = 0}^{\mu-1} e^{\complex a_n}\hat{\mathcal{T}}_n e^{-\frac{2\pi \complex nm}{\mu}}\,.
\end{equation}

Both the simple null model and the Fourier null model were used previously on time-series output of coupled Kuramoto oscillators, and they exhibit different dynamics from those of the coupled Kuramoto oscillators~\cite{Bassett2013,Bassett2014III}.


\subsection{Persistent homology applied to the Kuramoto model and null models}

We apply the WRCF to functional networks created from the output of two time regimes of the Kuramoto model, one time regime for the Kuramoto model, the simple null model, and the Fourier null model. We run the filtrations up to filtration step $1800$ for the first time regime and up to $2000$ for the second; we go up to filtration step $1100$ for cases in which we only consider one time regime. 
The total number of edges in the network, and thus the total number of possible filtration steps, is $8128$. The number of filtration steps thereby correspond to respective edge densities of 0.22, 0.25, and 0.14 for the three examples above; in each case, this amounts to a threshold that is approximately in the middle of the range of the edge-weight values. The Masters thesis of Stolz \cite{Stolz2014}, which is a precursor to the present paper, also applied PH to networks created from the Kuramoto model, and such an example was subsequently also studied using Betti curves by other authors\cite{Sizemore2015}. 

\begin{figure}[h!]
{Time regime I \ \ \ \ \ \ \ \ \ \ \ \ \ \  \ \ \ \ \  \ \ Time regime II}
 \\ \ \ \centering\includegraphics[width=0.209\textwidth]{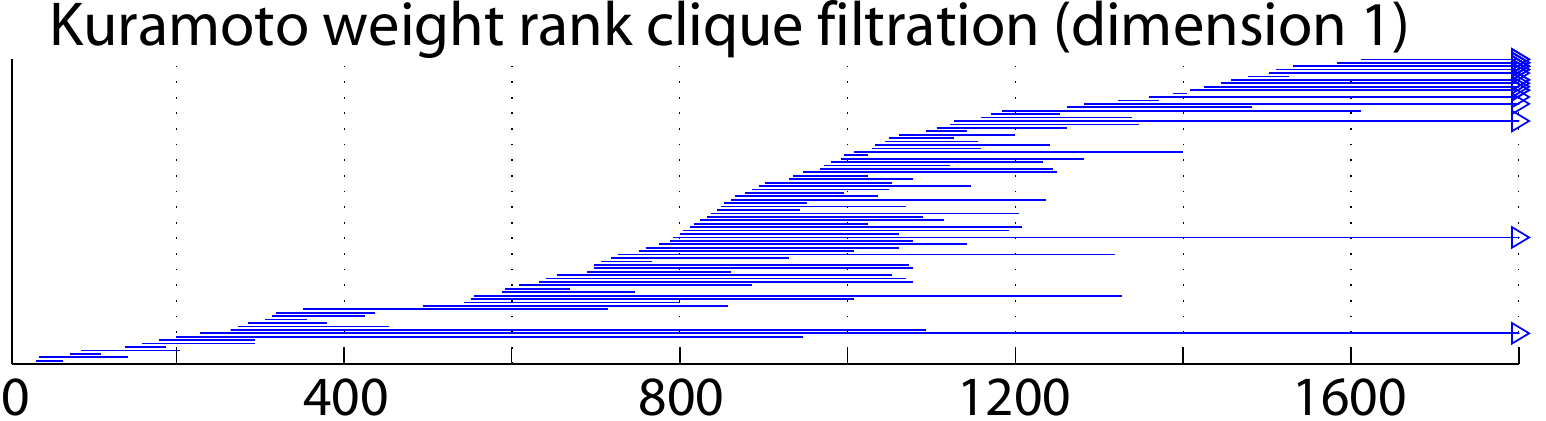} \ \ \ 
 \includegraphics[width=0.21\textwidth]{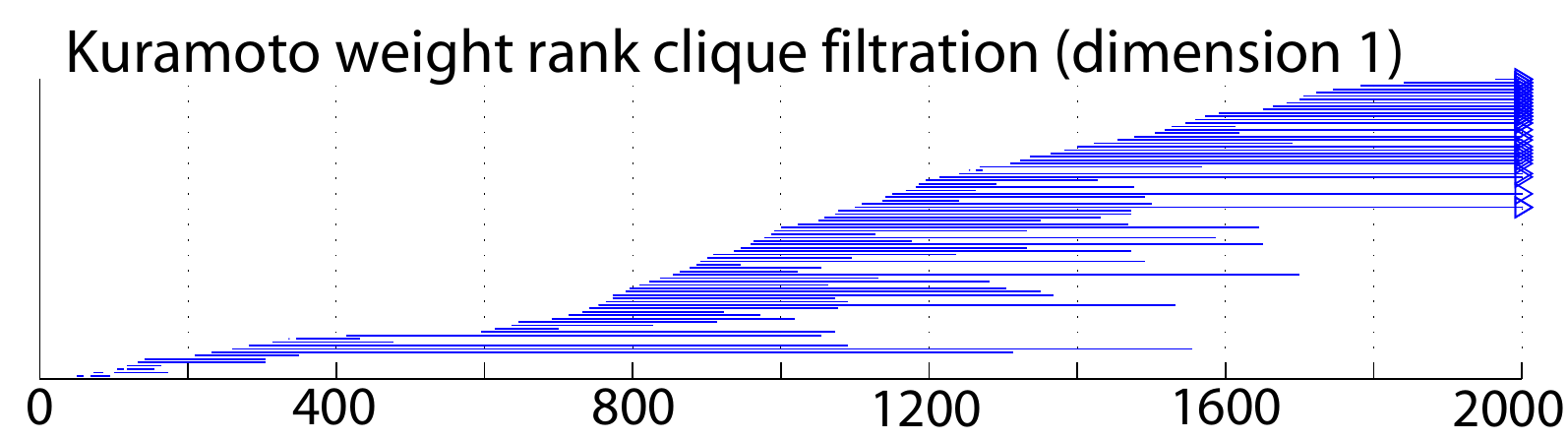}
\includegraphics[width=0.227\textwidth]{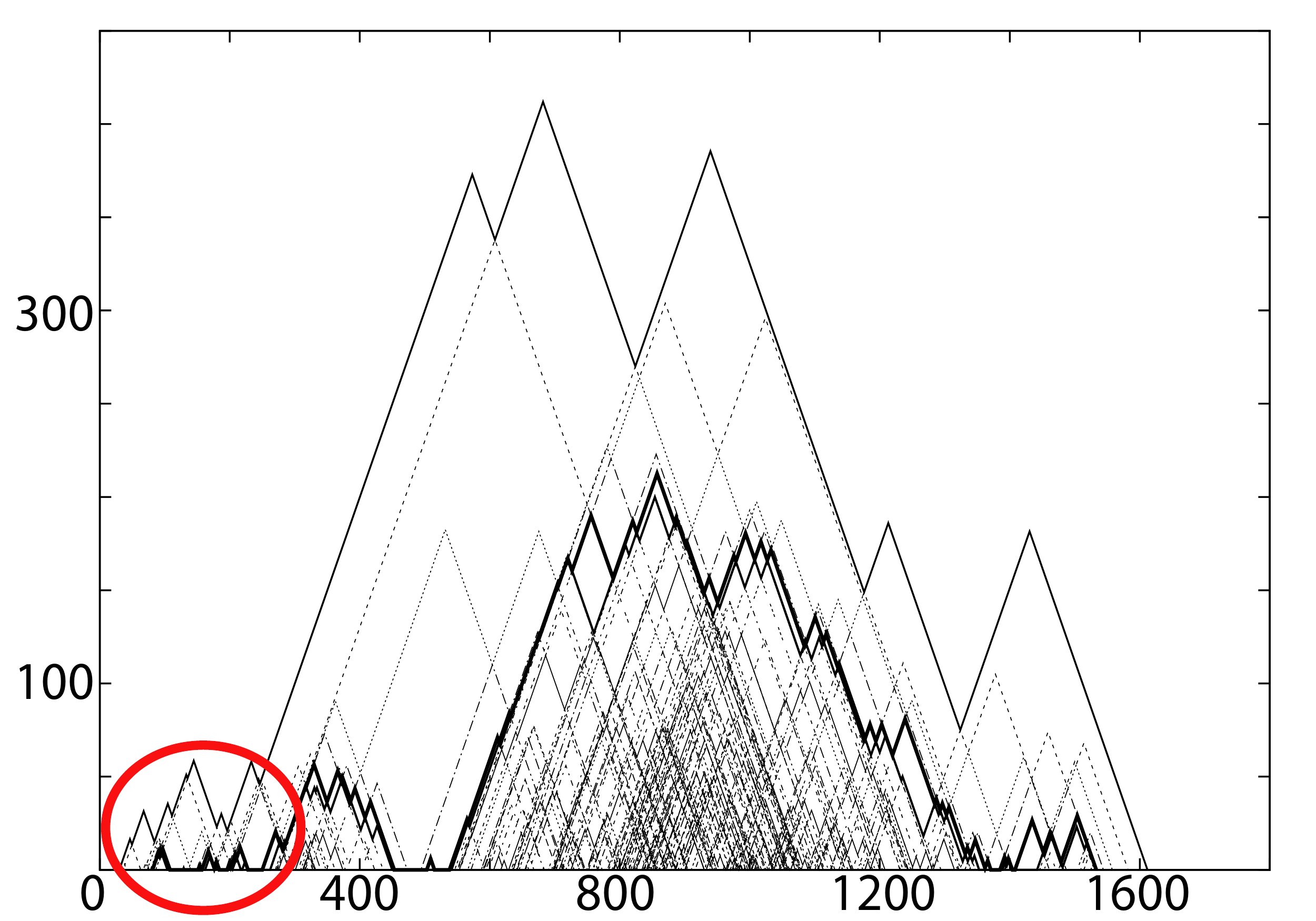}
\includegraphics[width=0.23\textwidth]{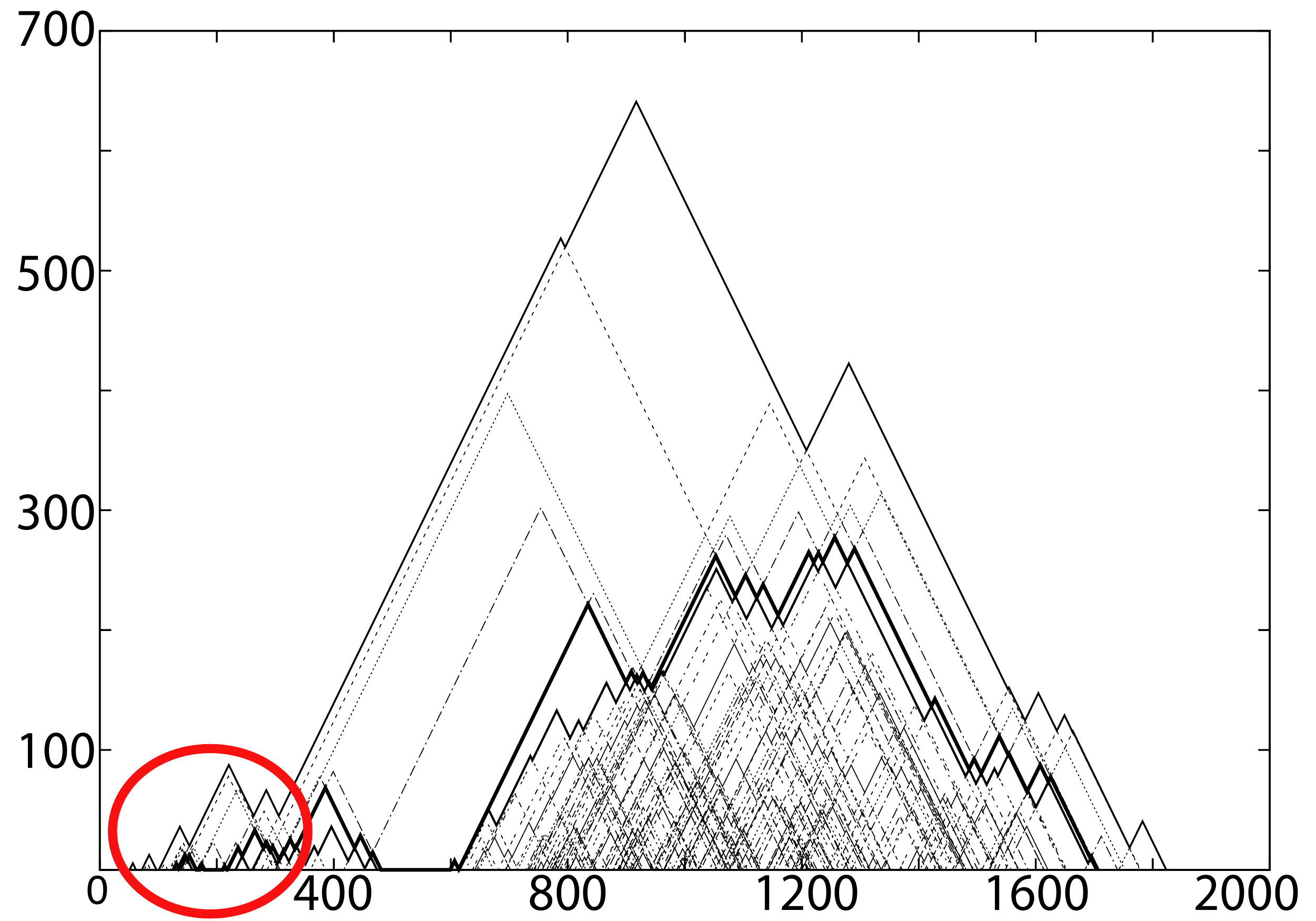} \\
\caption{Dimension-1 barcodes and persistence landscapes for the WRCF for the two time regimes (time steps 1--250 and time steps 251--500) of time-series output of the Kuramoto model. The horizontal axis represents the filtration steps in both the barcodes and the landscapes. The vertical axis in the persistence landscape captures the persistence of the features in the barcode. In the first row, we show the barcodes for dimension 1. In the second row, we show persistence landscapes (although we ignore infinitely-persisting bars in the barcodes). The short peaks at the beginning of the filtration in the persistence landscapes that are indicated by the red ellipses represent loops formed within communities. The most prominent difference between the two landscapes is the occurrence of high peaks in the second time regime; these peaks correspond to persistent loops in the network that are formed between communities.
}\label{fig:KuraPersistenceLandscapes}
\end{figure}

As we described in Section~\ref{sec:Intro}, we focus our analysis on topological features in dimension $1$, so examine loops in the network. In the first row of Fig.~\ref{fig:KuraPersistenceLandscapes}, we show the 1-dimensional barcodes for the networks constructed from time regime 1 (i.e., the first 250 time steps of the dynamics) and time regime 2 (i.e., time steps 251--500 of the dynamics) for the WRCF of the Kuramoto model. The barcode for each time regime includes several very short-lived bars between filtration steps 50 and 300. For the second time regime, we find more short bars for a longer filtration range at the beginning of the barcode. The $1$-loops that correspond to these short bars are all formed within the strongly synchronized communities. In fact, in time regime 1, the first 44 bars in the barcodes represent intra-community loops; in time regime, only 2 of the first 28 bars represent intra-community loops. As strong intra-community edges are added to the simplicial complexes, they start to cover the $1$-loops with triangles (i.e., $2$-simplices), and the loops disappear from the filtration. 

In the second row of Fig.~\ref{fig:KuraPersistenceLandscapes}, we show the persistence landscapes that we construct from the $1$-dimensional barcodes. We ignore infinitely-persisting bars in the barcode. (We also studied persistence landscapes including the infinite bars as features with a death time that corresponds to the maximum filtration value but did not obtain any additional insights that way.)

As expected, the landscapes have a group of small peaks early in the filtration for both time regimes. This feature occurs in a longer filtration range in the second time regime before more persistent loops appear. In the second time regime, some of the peaks that occur in the beginning of the filtration appear to almost double their heights to values of about $100$. In contrast, in the first time regime, peaks at a similar location are about half as high (i.e., they are less persistent). 

The persistence landscapes reveal more persistent $1$-loops in the second time regime (i.e., between time steps 251 and 500) than in the first (i.e., between time steps 1 and 250), and the second time regime also appears to reveal a clearer separation between the group of the very early short peaks and a second group of medium-sized peaks towards the end of the filtration. For this second group of medium-sized peaks, we observe a larger absolute increase in persistence in the second time regime than for the shorter peaks in the beginning of the filtration. These observations reflect the dynamics of the two time regimes in the Kuramoto model\cite{Bassett2013}. In time regime $1$, there is strong synchronization within the communities, and such dynamics are reflected by the appearance of short-lived intra-community $1$-loops (corresponding to the short peaks in the persistence landscapes) at the beginning of the filtration. In the second time regime, the amount of global synchronization is more prominent than in the first time regime. Moreover, in addition to intra-community loops, some of the peaks at the beginning of the filtration now represent inter-community loops, which are more persistent than the loops within communities.  Additionally, as some of the peaks that correspond to inter-community loops have shifted to the beginning of the filtration, there is an increase in the gap between the initial group of peaks and the group of medium-sized peaks at the end of the filtration. In general, we observe an increase in the persistence of the peaks in the landscapes due to the stronger synchronization between the communities. These observations are much easier to visualize using persistence landscapes than using barcodes. 
 
We calculate pairwise $L^2$-distances between all dimension-1 persistence landscapes, and we note that $L^2$ distance has been used previously to compare persistence landscapes in an application to protein binding \cite{Kovacev2015}. The $L^2$ distance between the two time regimes is 27078. Given the length of the support of the landscapes and the function values that they attain, this is a large distance, which captures the aforementioned visible differences between the landscapes. The $L^2$ distance is unable to capture the fact that the peaks that appear early in the filtration in the first time regime correspond to loops between nodes within one community, whereas they correspond to loops that form between nodes of different communities in the second time regime. Consequently, this feature does not contribute to the value of the distance.

\begin{figure}[h!]
 \ \includegraphics[width=0.1525\textwidth]{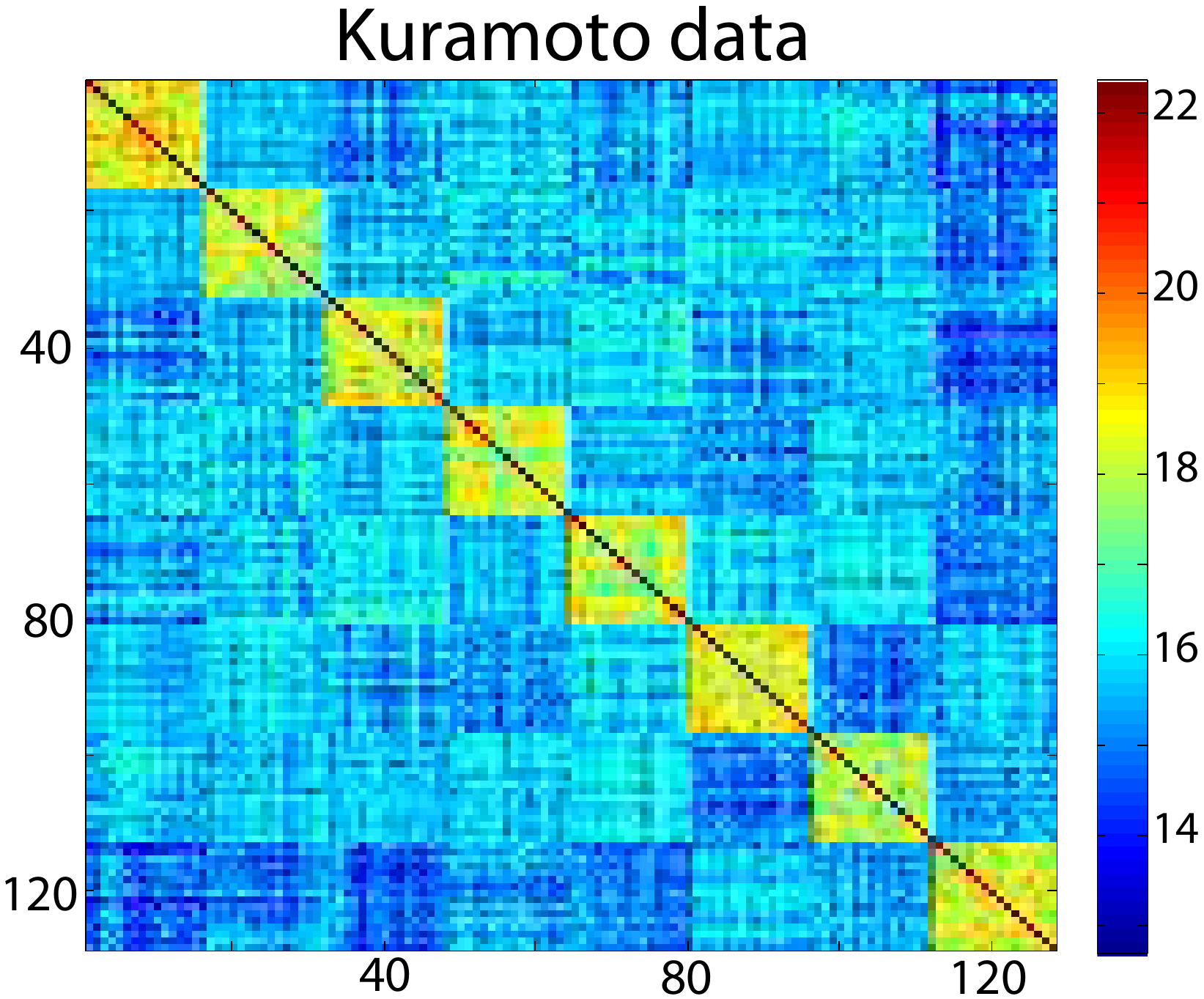}\ 
 \includegraphics[width=0.152\textwidth]{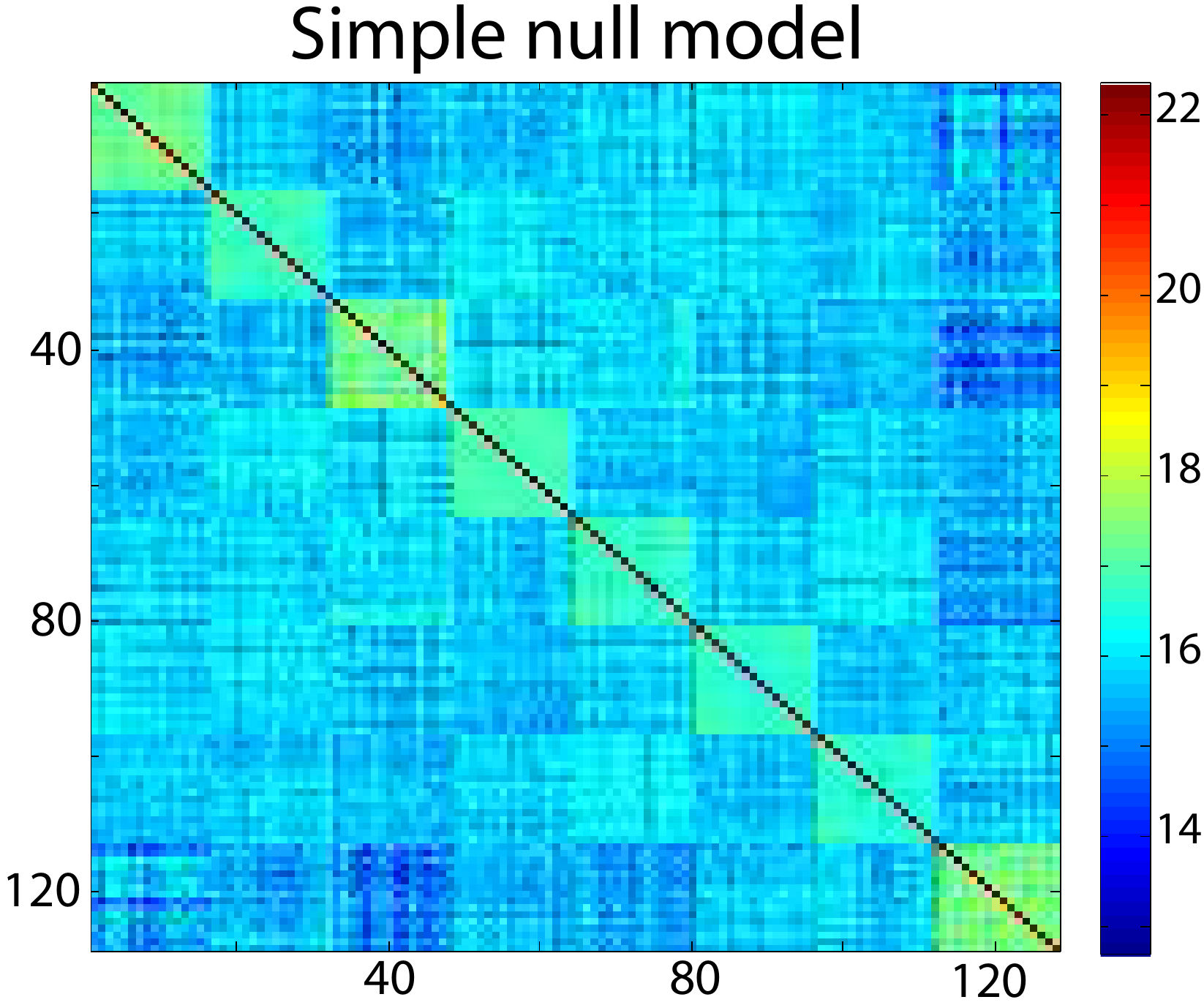}
 \includegraphics[width=0.1515\textwidth]{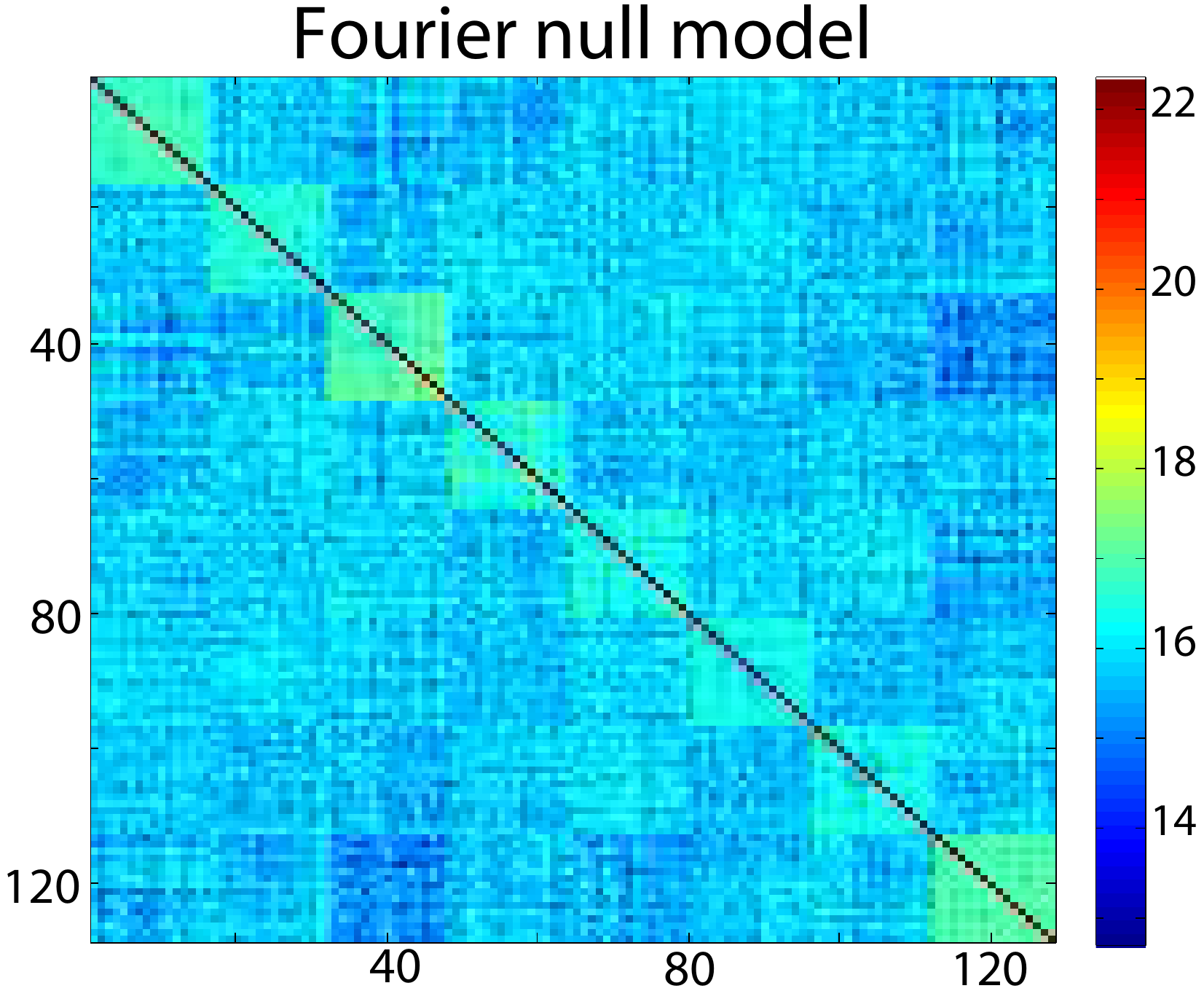}\\
\ \includegraphics[width=0.152\textwidth]{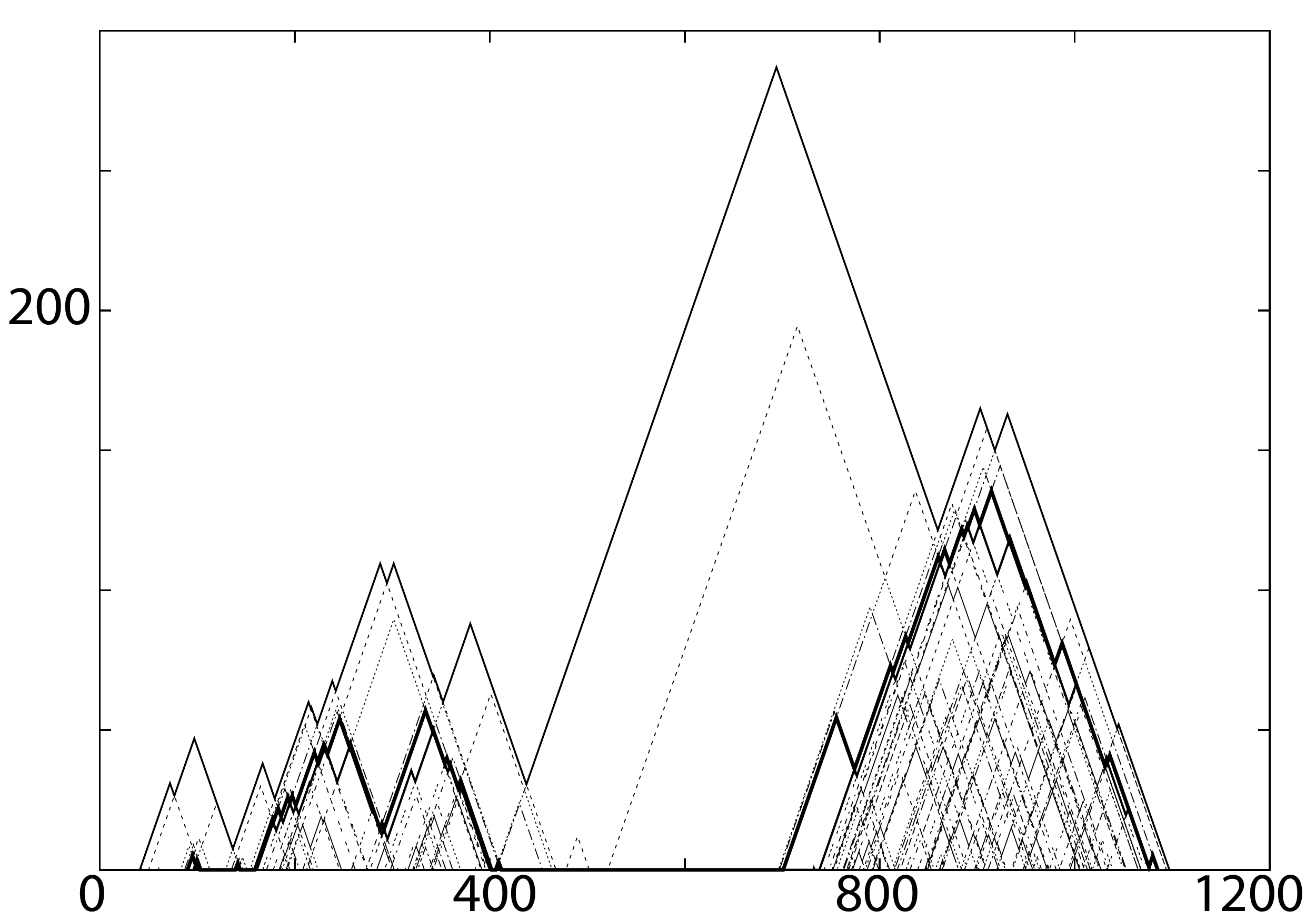}
 \includegraphics[width=0.152\textwidth]{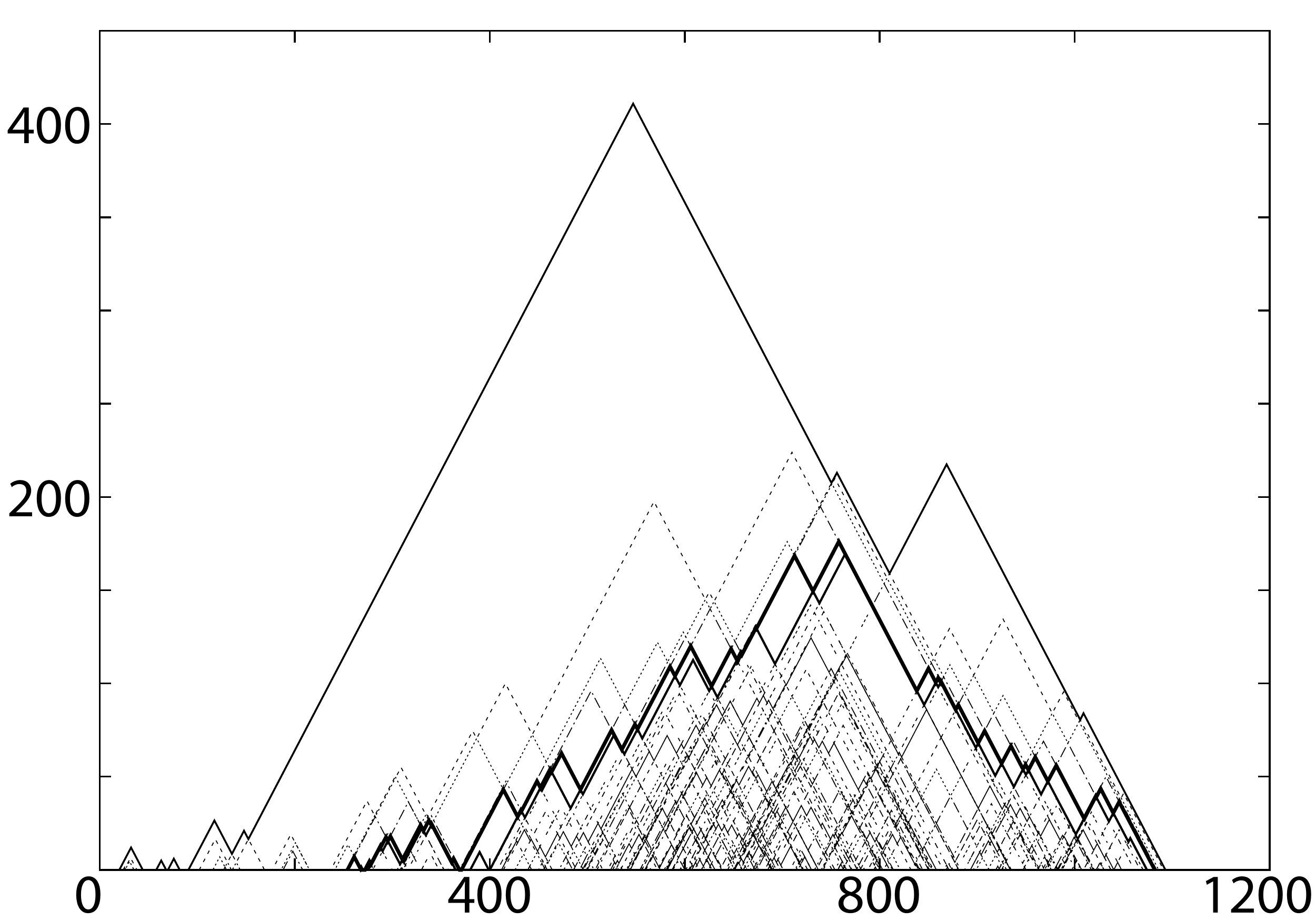}
 \includegraphics[width=0.152\textwidth]{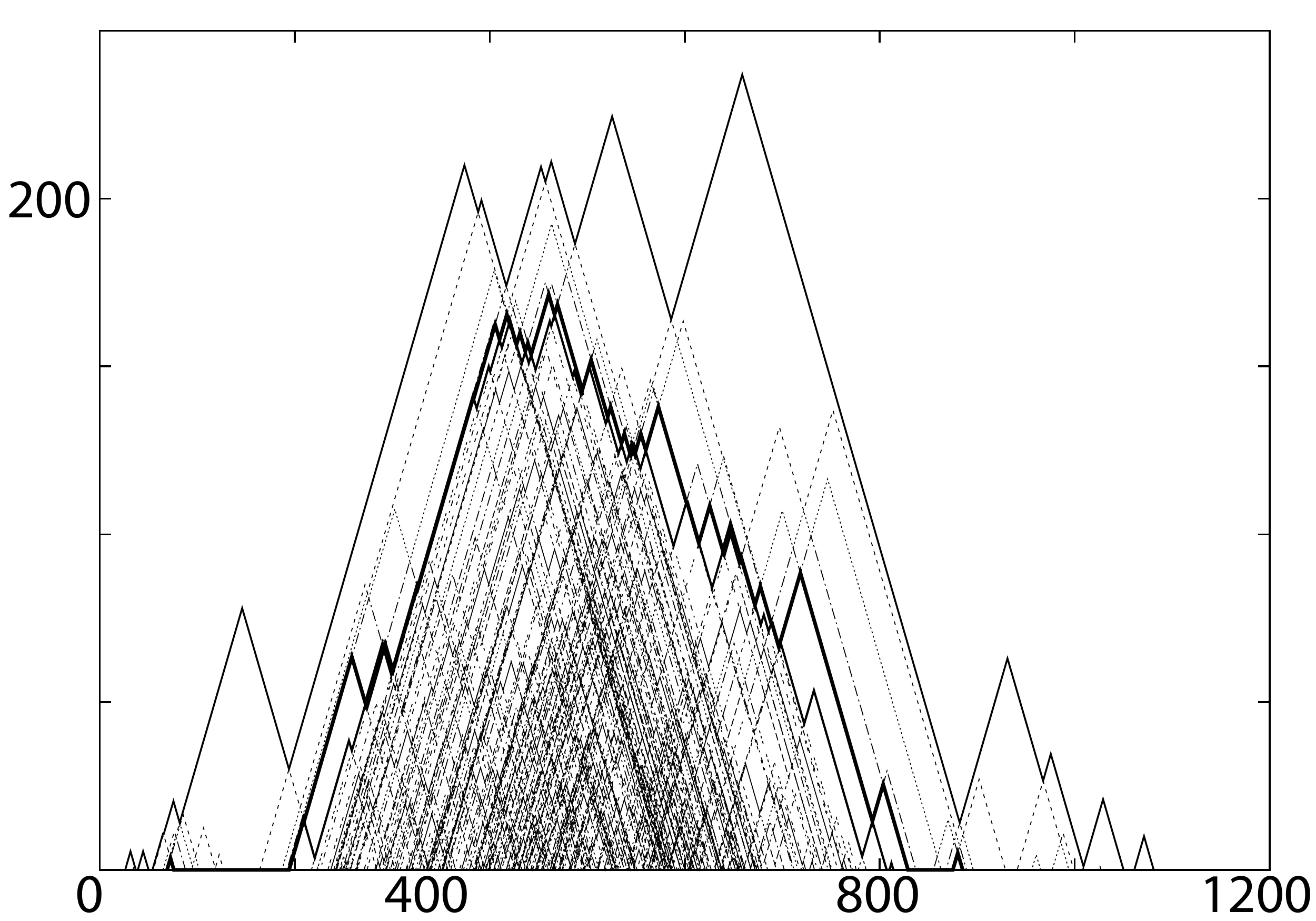}\\
 \caption{(Top row) Functional networks for (left) the Kuramoto model, (center) the simple null model, and (right) the Fourier null model. (Bottom row) Dimension-1 persistence landscapes for the WRCF of (left) the Kuramoto model, (center) the simple null model, and (right) the Fourier null model using one time regime and ignoring infinitely-persisting bars. The persistence landscapes illustrate differences in the occurrence of loops in the three different networks. Most prominently, these differences manifest in the heights and distributions of the peaks in the landscapes, which appear to exhibit a stronger separation along the filtration between groups of peaks of different heights for the Kuramoto model than in the two null models.} 
 \label{Fig:Kuramotonull models}
 \end{figure}

We also compare the Kuramoto model to the two null models that we discussed in Section \ref{sec:KuramotoNull}. To do this, we construct a functional network by considering a single time regime that consists of 500 time steps. In Fig.~\ref{Fig:Kuramotonull models}, we show the weighted adjacency matrices of the three functional networks, and we also show their corresponding persistence landscapes based on WRCFs of the functional networks. One can observe clearly that there is stronger intra-community synchronization for the Kuramoto times series than for the null models, as there is a very distinct group of short peaks at the beginning of the filtration (which, as we discussed above, is also the case for the Kuramoto model when performing separate calculations in the two time regimes).

Again, the corresponding loops occur within communities. The peaks in the Kuramoto landscape appear to be separated from a second group of short peaks further along in the filtration. Between the two groups of peaks, there are two strikingly higher peaks that correspond to persistent loops, which appear to be formed by connections between different communities. For both null models, we also observe groups of short peaks at the beginning of the filtration, but these are less persistent and less clearly separated from other peaks than for the Kuramoto model. Indeed, we do not see any separation at all for the Fourier null model, which exhibits a much weaker intra-community synchronization than the simple null model. Moreover, the persistence landscape for the Fourier null model appears to be ``noisier," as the majority of the peaks in the landscape have similar persistences and appear in similar areas of the filtration. 

The peaks in the landscapes of the null models appear to have a very different distribution along the filtration than is the case for the Kuramoto model. They also possess more medium-sized and long persisting features than we observe in the Kuramoto data. These features occur in parts of the filtration in which the Kuramoto data has a smaller number of peaks. They consist of inter-community loops and are a symptom of the weaker intra-community and stronger inter-community synchronization. The null models thus appear to have more topological features in the form of loops than is the case for the Kuramoto data. This is consistent with previous observations of null models in other studies~\cite{Petri2013,Giusti2015,Sizemore2016}. The fact that there are fewer persistent loops in the Kuramoto model than in the null models implies that there are more high-dimensional simplices (e.g., triangles and tetrahedra) in the corresponding network than in the networks constructed from the null models.

To distinguish between the three landscapes, we calculate the $L^2$ distances between them. The $L^2$ distance between the Kuramoto landscape and the Fourier null-model landscape is 13540 the $L^2$ distance between the two null-model landscapes is 13263, and the $L^2$ distance between the Kuramoto landscape and the simple null-model landscape is 11703. Again considering the support of the landscapes and the attained function values, we see that three distances can be construed as large.

For the Kuramoto model, we find that PH can detect the dynamics of the system and that the persistent landscapes are rather different for the Kuramoto model and the null models. The $L^2$ distances between landscapes underscore these differences. We are also able to distinguish between the two null models using persistence landscapes. In contrast to conventional wisdom\cite{Ghrist2008,Carlsson2009}, we do not find for our examples that only the persistence of topological features distinguishes between signal and noise. In fact, the short bars at the beginning of the filtration of the Kuramoto model carry important information about the dynamics, and the medium-sized persistent peaks in the Fourier null model are a symptom of the weaker intra-community and stronger inter-community synchronization in that model. We therefore assert that the position of features in the barcode is as important as persistence length for their interpretation in our examples, and this provides an important point to consider for future studies. Note that persistent landscapes alone do not provide enough information to assess system dynamics. It is only by combining them with information about nodes that forming loops (which are represented by certain groups of peaks) that we are able to obtain conclusions about intra-community and inter-community synchronization.


\section{Example II: Task-based fMRI data} \label{sec:Ex2}

\subsection{Human brain networks during learning of a simple motor task}

We use a data set of functional brain networks from experiments that were first analyzed by Bassett et al. \cite{Bassett2011}. The data set was collected to study human subjects during learning of a simple motor task, and a full description of the experiments conducted is available in \onlinecite{Bassett2011}. We apply a WRCF to functional networks, and we compare our findings to previous studies on these and similar networks~\cite{Bassett2011,Bassett2013cp,Bassett2014}. The functional networks are based on functional magnetic resonance imaging (fMRI) time series\footnote{See [\onlinecite{eklund2016}] for a recent discussion of fMRI inferences and potential perils in the statistical methods in use in neuroimaging.} from 20 healthy subjects who undertook a motor-learning task on three days (during a five-day period). During the imaging of the subjects, an ``atlas'' of $112$ brain areas was monitored while they were performing a simple motor-learning task (similar to a musical sequence), which they executed using four fingers of their non-dominant hand. For each subject and for each day of the study, the fMRI images are interpreted as $2000$ time points for each monitored brain region. The brain regions and their time series were used subsequently to construct functional networks based on a functional connectivity measure known as the coherence of the wavelet scale-2 coefficients. This measure was applied to the time series to determine edge weights between every pair of brain regions in the network. The weighted adjacency matrices for the functional networks were then corrected for a false-discovery rate, as matrix elements under a certain threshold (which represents a correlation amount that one expects to occur at random) were set to $0$. The other matrix elements were retained.

The functional networks that we just described were studied previously using community detection by Bassett et al. \cite{Bassett2011}, whose results suggest that there is a significant segregation of the nodes in the functional networks into a small number of different communities with densely-weighted connections inside the communities and sparsely-weighted connections to nodes in other communities. Within these communities, certain nodes appeared to remain in the same community during the experiment, whereas others (the ``flexible'' ones) often switched between different communities. 

There have also been studies of networks from a similar experiment but with medium-term learning and including training sessions \cite{Bassett2013cp,Bassett2014}. These networks have a noticeable core--periphery organization, with the sensimotor and visual regions of the brain grouped into a temporally ``stiff'' core of nodes, whose community memberships (in contrast to flexible, peripheral nodes) do not change much over the course of the learning task\cite{Bassett2013cp}. It was also shown subsequently that the interaction between primary and secondary sensorimotor regions and the primary visual cortex decreases as the regions (presumably) become more autonomous with task practice 
\cite{Bassett2014}. 

Because we observed short-lived 1-dimensional loops in the beginning of the filtrations for the Kuramoto model in a simulated setting with community structure in oscillator connectivity, we will explore whether the fMRI data exhibits similar features during the three observation days.


\subsection{Persistent homology applied to the task-based fMRI data}

\begin{figure}[h!]
\includegraphics[width=0.48\textwidth]{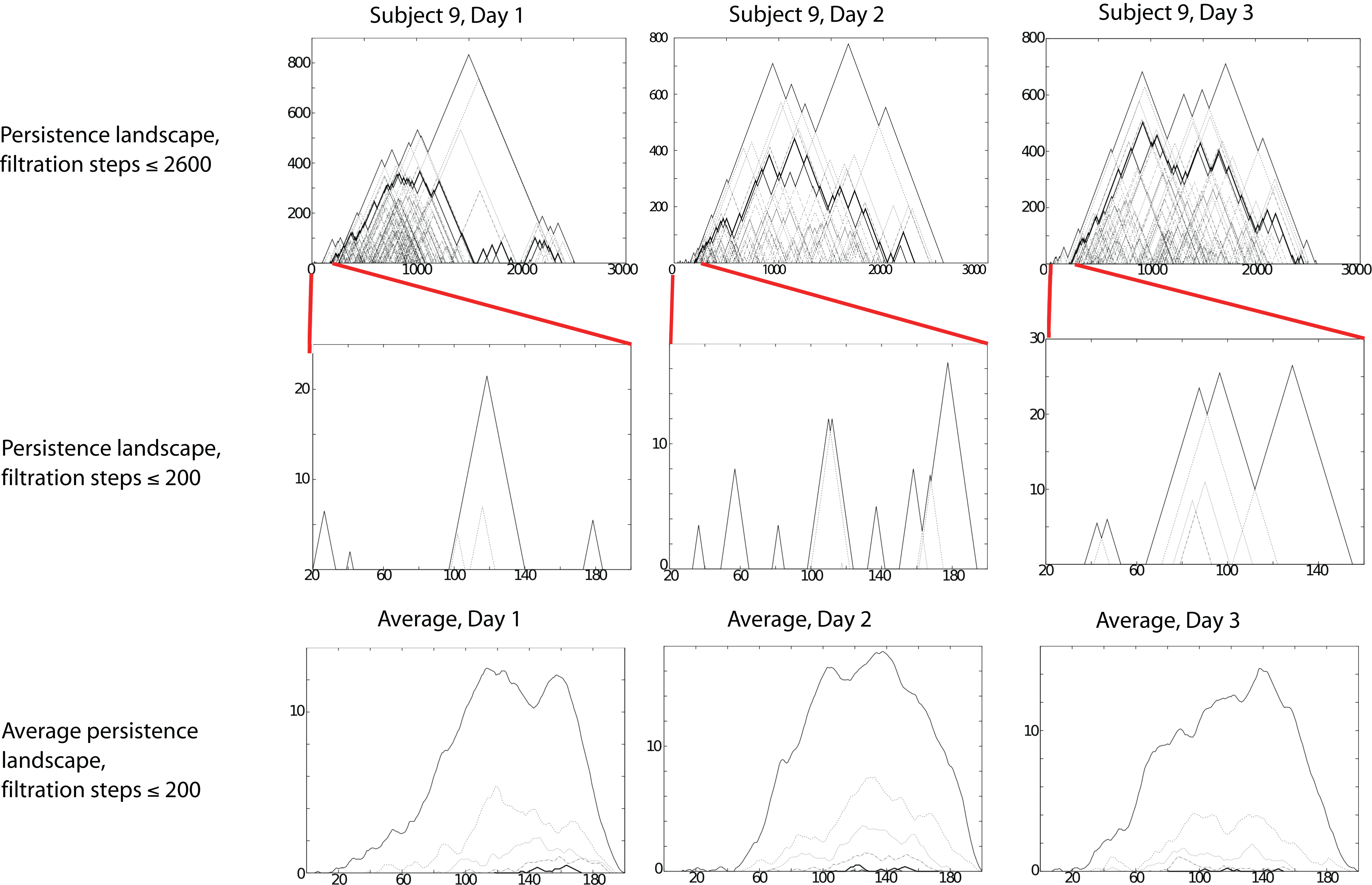}
\caption{Persistence landscapes for dimension 1 of the WRCF applied to the human brain networks. (First row) Persistence landscapes for subject 9 based on filtration steps 1--2600 for days 1, 2, and 3. (Second row) Persistence landscapes for subject 9 based on filtration steps 1--200 for days 1, 2, and 3. (Third row) Average persistence landscapes over all subjects for days 1, 2, and 3. We observe on average that short peaks occur in the first 200 filtration steps of the landscapes.
}\label{Tab:PersLandscfMRI}
\end{figure}

We run the WRCF until filtration step 2600, which is when $42 \%$ of the edges are present in the network. (Note that using more filtration steps leads to very long computational times.) We again focus our analysis on topological features in dimension $1$. 
We construct persistence landscapes for dimension 1 (omitting infinitely persisting $1$-loops). In Table~\ref{Tab:PersLandscfMRI}, we summarize our results for one particular subject and for the whole data set. We use this subject to illustrate a representative example of the particular landscape features that we observe in the data.

Similar to the Kuramoto oscillators in Section \ref{sec:Ex1}, we find a group of small peaks at the beginning of the filtration (between filtration steps 1 and 200). We can see this group very clearly both by magnifying either the landscape of individual subjects or the average landscape, where the height of the peaks is only slightly smaller than for the peaks in the individual landscape that we show. This feature of the heights indicates that a group of short peaks arises in the beginning of the filtration in the majority of the barcodes. We also consider the standard deviation from the average landscapes in the first 200 filtration steps. For all three days, it is very small: it is 127 for the first day, 167 for the second day, and 126 for the third day.

We expect the observed short peaks in the beginning of the filtration to be associated with network communities, which have been observed previously using other methods\cite{Bassett2011}. We observe, in particular, that these short peaks undergo changes on day 2: during filtration steps 20 to 60, some of the peaks that are present in the landscapes for days 1 and 3 vanish, and more persistent peaks occur for day 3 than on the other two days between filtration step 80 and 200. This appears to suggest that there is a change in community structure that takes place on day 2, with either (1) very strong synchronization in some of the communities, leading to very short-lived $1$-loops; or (2) very strong individual differences between the subjects, leading to the vanishing of peaks in the average landscapes for the first 50 filtration steps. The particularly persistent peaks on day 2 could represent either persistent loops between different communities or loops that occur due to sparse intra-community connections.

We calculate pairwise $L^2$-distances between all dimension-1 persistence landscapes. We create distance vectors, which we use as an input for $k$-means clustering and average linkage clustering for $k = 3$, and we obtain the same qualitative result for both methods. We find that $9$ of the $20$ distance vectors that correspond to persistence landscapes from day $1$ are assigned to a common group (together with a small number of landscapes from days 2 and 3), whereas $11$ and $10$ landscapes from days 2 and 3, respectively, are assigned together to a separate group. We summarize our results in Table \ref{Tab:ThreeDayDistances}.

\begin{table}
\begin{tabular}{p{1.7cm} | c | c | c }
 & \footnotesize{Cluster 1} & \footnotesize{Cluster 2} & \footnotesize{Cluster 3}\\
  \hline
   \hline
 Day 1 & {9}    &  6   &  5 \\
 \hline
  Day 2 & 5    &  4   &  {11} \\
   \hline
   Day 3 &   5    &  5   &  {10}\\
    \hline
\end{tabular}
\caption{Results for $k$-means clustering and average linkage clustering of pairwise $L^2$-distance vectors of persistence landscapes for $k = 3$. 
}\label{Tab:ThreeDayDistances}
\end{table}

\begin{figure}[h!]
\includegraphics[width=0.5\textwidth]{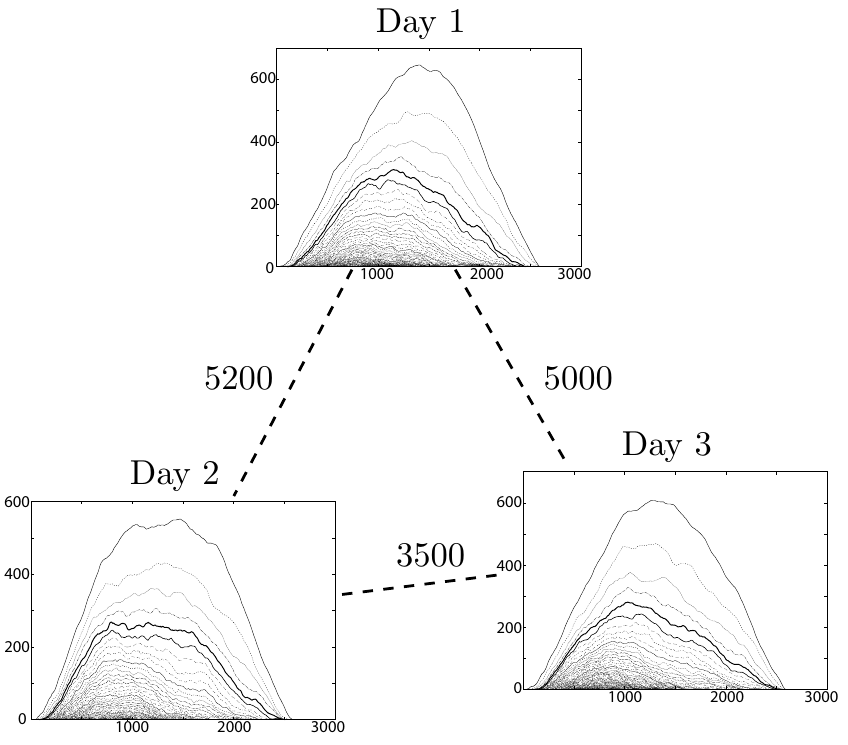}
\caption{Visualization of average persistence landscapes for days 1, 2, and 3 of task-based fMRI networks. The distance between the landscape for day $1$ and the other two landscapes is larger than that between the landscapes for days 2 and 3. (The $L^2$ distances between them are 5200 between days 1 and 2, 5000 between days 1 and 3, and 3500 between days 2 and 3.) The standard deviations from the average landscapes are larger than the calculated distances, so these values need to be interpreted cautiously. We also observe a shift to the left of the landscape peak during the three days, indicating that the particularly persistent $1$-loops in these networks arise earlier in the filtration for the later days. In other words, they are formed by edges with a higher edge weight, indicating that there is stronger synchronization between the associated brain regions. 
}\label{Fig:AverageLanscapesfMRI}
\end{figure}

We also consider the average dimension-1 landscapes for WRCF steps $1$--$2600$ and calculate the $L^2$-distances between them. We show the results of these calculations in Fig.~\ref{Fig:AverageLanscapesfMRI}.

The distances between the average landscape for day 1 and the subsequent days of the experiment indicate that the WRCFs on average are able to detect changes in the functional networks across the filtration range. Based on the distances, we observe that most of these changes occur between the first and the second day. However, the standard deviations from the average landscapes are a factor of about $4$ larger than the distances between the landscapes, and one therefore needs to be cautious about interpreting the results of these calculations. In a permutation test with 10000 regroupings of the landscapes, we do not find the distances to be statistically significant. We obtain $p$-values of about $0.4$ for the distance between the average landscapes of day 1 and day 2, about $0.85$ for the distance between the average landscapes of day 2 and day 3, and about $0.6$ for the distance between the average landscapes of day 1 and day 3.

For the average landscapes in Fig.~\ref{Fig:AverageLanscapesfMRI}, we also find that that the primary peak of the average landscapes shifts to the left over the course of the three days. This implies that the edge weights (between the brain regions) that give rise to persistent $1$-loops increase on average over the three days (presumably due to stronger synchronization). This can either mean that loops present on the first day synchronize more on the second and third day, or that new loops that appear on days $2$ and $3$ consist of more synchronized edges. Brain regions that synchronize in a $1$-loop in a network may be an indication of an interesting neurobiological communication pattern that in this case also becomes stronger over the course of the learning process. To analyze the most frequently occurring edges involved in these loops, we extract ``representatives'' for all loops in dimension $1$ across all subjects and days. (See  Fig.~\ref{Fig:Loop} for an illustration of two different representatives of a loop in a network.) For each day, we construct a network, which we call the ``occurrence network,'' using the same nodes (i.e., brain regions) that we used before and assign every edge an edge weight that is equal to the number of occurrences of that edge in $1$-dimensional loops in the subjects on the given day. We then perform a WRCF on the three occurrence networks and study representative loops given by the algorithm. In Table~\ref{table: Brain regions} in Appendix \ref{appB}, we list the brain regions that we find in loops that consist of edges that occur at least $50$ times in functional networks in the subjects.
We now examine loops in the occurrence networks. These particular loops may not correspond exactly to loops in the functional networks. For example, individual edges with high edge weights that are part of a loop in
the occurrence network may be part of a variety of different loops in the functional networks, rather than part of one specific loop
that occurs in many of the functional networks. Nevertheless, it is very likely that such loops are also loops in the functional network. 
One also needs to consider that the representative loops given by the software {\sc javaPlex} are not necessarily chosen optimally or ``geometrically nice''\footnote{For example, a loop may be represented by a double loop.} representatives of the loop~\cite{Adams2014}. (See Fig.~\ref{Fig:Loop} for an illustration of different representatives of the same loop.)
We address the issue of the algorithm's choice of representatives to some extent by using PH on the occurrence network, but even then we cannot rule out possible artifacts. There exist loops in the occurrence networks that remain stable across the three days, although other loops occur on only one or two days. There also seem to be more loops that occur at least 50 times in the functional networks on days $2$ and $3$ than on day $1$. It would be useful to study the brain regions involved in the listed loops (see Table~\ref{table: Brain regions} in Appendix \ref{appB}) to investigate their biological role in motor-learning tasks.

\begin{figure}[h!]
\includegraphics[width=0.5\textwidth]{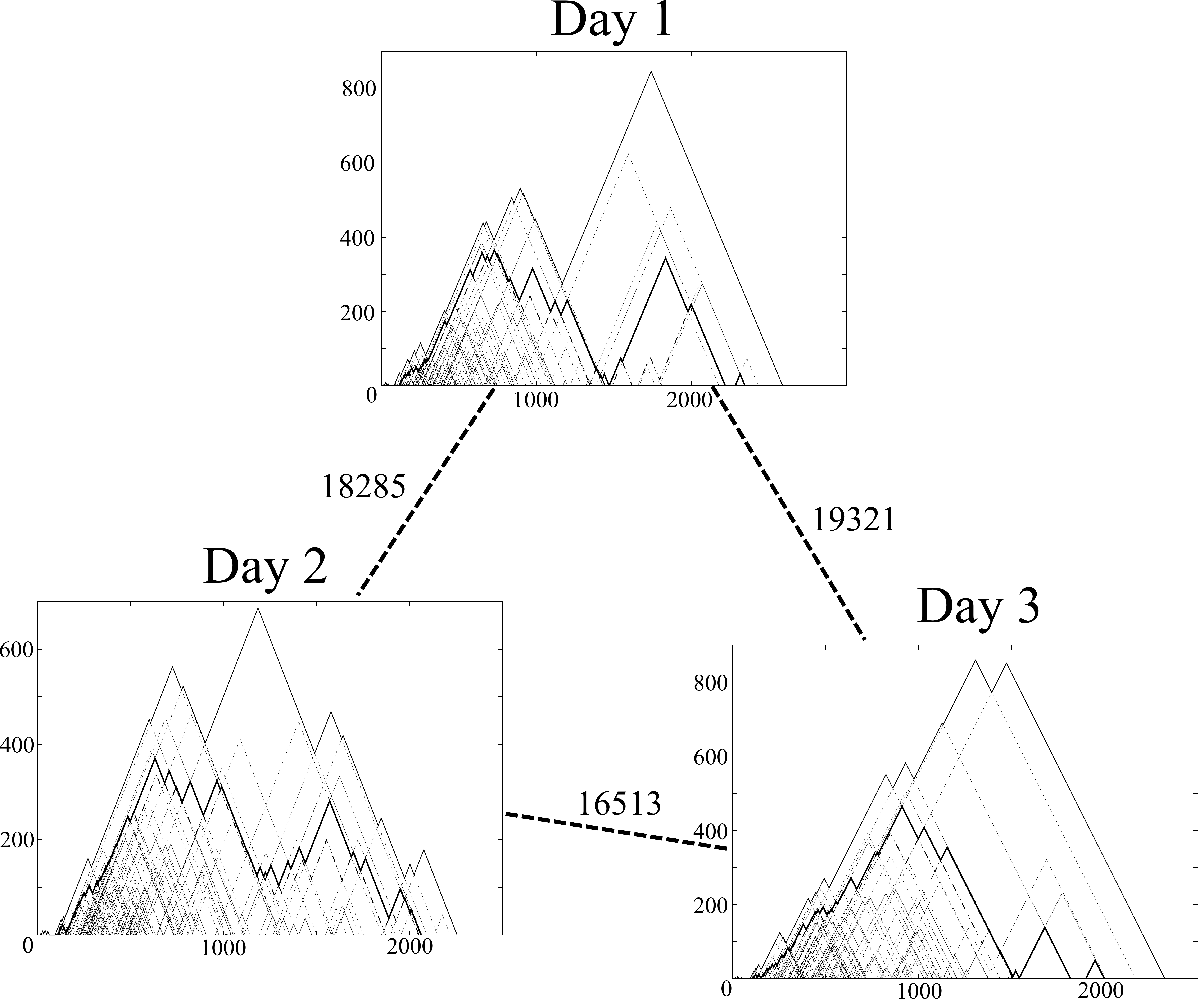}
\caption{Visualization of persistence landscapes based on average functional networks on days 1, 2, and 3 of the motor-learning task. The distance between the landscape for day $1$ and the other two landscapes is larger than that between the landscapes for days 2 and 3. (The $L^2$ distances between them are 18285 between the first and second days, 16513 between the first and third days, and 19321 between the second and third days.) We find short peaks at the beginning of the filtration for all three landscapes, and larger peaks begin earlier in the filtration on day 3 than on day 1.
}\label{Fig:PersLandscAveragefMRI}
\end{figure}

Finally, we also apply WRCF to the average networks for each of the three days. To create the average networks, we take the mean of the edge-weight values over all 20 subjects for each day separately and study the resulting network. We show the corresponding landscapes in Fig.~\ref{Fig:PersLandscAveragefMRI}.

As with the average landscapes, we find that the landscapes for the average networks have very short peaks in the beginning of the filtration. There are more persistent features (e.g., larger peaks) on day 1 and day 3 than on day 2, and we even find (as in the average landscapes) that the larger peaks appear earlier (at about filtration step 400) in the filtration on day 3 than on day 1 (where they appear at about step 900). Additionally, on day 2, we observe many short peaks, especially in the later stages of the filtration. This is not the case for day 1 and day 3, so the day-2 landscape is strikingly different visually from the other two landscapes. When calculating $L^2$ distances, we again find that the landscape distance between days 1 and 2 and that between days 1 and 3 are larger than the landscape distance between days 2 and 3. From visual inspection, we see that this arises from the fact that the day-1 landscape appears to have a clearer separation of short and high peaks than the landscapes for the later days. Taken together, the results for the landscapes of the average networks mirror our prior results for the average landscapes.


\section{Conclusion and discussion} \label{sec:ExConcDisc}

We have illustrated applications of persistent homology to functional networks constructed from time-series output of the Kuramoto model, null models constructed from the Kuramoto time series, and task-based fMRI data from human subjects. In all cases, we observed that non-persistent $1$-loops occur at the beginning of the filtrations. Although such non-persistent features are commonly construed as noise in topological data analysis \cite{Ghrist2008,Carlsson2009}, we observed that these features appear to be consistent with prior segregations of the studied networks into communities of densely-connected nodes. In one case (the Fourier null model), we even found that particularly persistent features appear to be linked to a network with a weak intra-community synchronization. These very persistent features in the null model, thus may represent noise. In other studies of PH using (different) null models~\cite{Petri2013,Giusti2015,Sizemore2016}, it was also observed that the null models often exhibit a richer topological structure than the empirical data. One could thus perhaps interpret the persistent features in the Fourier null model as features of the null model rather than as noise. Our results on the importance of non-persistent features match previous observations for synthetic examples with barcodes that consist of short intervals (which are commonly be construed as noise), but the differences between the corresponding persistence landscapes for the various spaces are nevertheless statistically significant~\cite{BubenikPC}. Our results are also consistent with the findings of a study on protein structure using PH for which bars of any length in the barcodes were equally important~\cite{Xia2014}. For weighted networks, we suggest that when using a filtration based on edge weights, one needs to consider the actual birth and death times of filtration features (such as $1$-loops) in addition to their persistence to be able to determine whether they should be construed as part of noise or part of a signal. In particular, in the present paper, we observed that the early appearance of $1$-loops in a filtration are important distinguishing features of these data. They may also yield important insights on the geometry~\cite{BubenikPC} of data\footnote{Note that we use the term ``geometry'' for properties that are called ``shape'' in other contexts (see, e.g.,~[\onlinecite{macpherson2012}]) to avoid confusion with our previous usage of the term ``shape.''}.

We also found --- both by calculating average persistence landscapes and studying landscapes of average networks --- that persistence landscapes for dimension $1$ of the weight rank clique filtration (WRCF) are able to capture changes in the studied functional brain networks during the process of learning a simple motor task. Because we did not consider infinitely-persisting features and only included filtration steps 1--2600 when creating the landscapes, our result also suggests that the medium-lived (when compared to the the full filtration length) persistent $1$-loops are able to capture changes in the network, so it is not always necessary to consider a full WRCF to study the dynamics of a system. This observation is similar to a finding in Bendich et al. \cite{Bendich2014}, who observed in their study that medium-scale barcode features were able to distinguish human brain artery networks from different age groups. This again suggests that persistence length should not be the only measure of signal versus noise when applying PH. We also found that the persistent features that dominate the middle part of the filtrations appear in earlier filtration steps on days 2 and 3 of the experiment than they do on day 1, which suggests that interesting dynamics in synchronization patterns are captured by medium-lived bars in the middle of a barcode.

As in other biological contexts, where PH has been applied successfully and has lead to new insights\cite{Petri2014,Dlotko2016,Curto2008, Dabaghian2012, Giusti2015,Bendich2014}, we find that PH 
can lead to fascinating insights about the dynamics of a system. 
We were able not only to detect symptoms of previously observed community segregation, but we also found notable differences between a setup with strong community structure (in the coupled Kuramoto oscillators) and weakly synchronized communities (in the associated null models). For the task-based fMRI data, we found that we can detect symptoms of community structure over the three days (in the short peaks at the beginning of the landscapes) of the data as well as changes in the $1$-dimensional loops that appear on average in the functional networks. On average, most of these changes appear to take place on the second day of the learning task. In particular, brain regions that yield $1$-loops in the functional networks on days $2$ and $3$ seem to exhibit stronger synchronization on average than those that yield $1$-loops on day 1. We obtained this observation both by calculating average persistent landscapes of the WRCF performed on individual functional networks and by calculating persistent landscapes based on the WRCF performed on average networks for each day. Although the landscape distances between the average landscapes are not statistically significant, our similar observations in both of our approaches suggest that our observations indeed reflect the average dynamics of the system. Our findings on $1$-dimensional loops thereby provide novel insights that complement previous studies of synchronization in functional brain networks. It would be desirable to repeat our study using larger data sets.
 
There is a known relation between homology and graph Laplacians \cite{chung1997}, and an interesting possible direction for future research would be to study possible connections between graph Laplacians (and, more generally, spectral graph theory) and our results on barcodes and persistence landscapes.

Using methods from topological data analysis for studying networks has the important benefit of being both mathematically principled and generalizable. However, for biological interpretation, it is necessary to include information on the specific nodes that are part of the topological features such as loops. Moreover, the interpretation of the results and importance of persistence versus position of a topological feature in the barcode can differ depending on which type of filtration is employed. Different topological features can also have different levels of relevance for different dynamical systems. For example, the occurrence of many medium-sized persistent features in the persistence landscape for the Fourier null model is a symptom of the weak synchronization in the communities, whereas the medium-sized persistent bars capture increasing synchronization in $1$-loops for the task-based fMRI data. It would be interesting to apply WRCF (and other types of filtrations) to different synthetic networks with underlying communities (e.g., using stochastic block models) to investigate such ideas further. Importantly, one should include both the persistence and the position of topological features in analysis of PH. It would also be beneficial to combine topological tools with additional methods, such as persistence images\cite{Adams2015}, to determine the exact topological features that are responsible for the detected differences between the persistence landscapes of the different networks.

In conclusion, we have shown that persistent homology and persistence landscapes can be applied successfully to functional networks (from either experimental data or time-series output of models), and that they can lead to fascinating insights, such as segregation of a network into communities and changes of network structure over time.


\section{Acknowledgements} \label{sec:Acknowledgements}

The experimental data was collected originally by Nicholas F. Wymbs and Scott T. Grafton through funding from Public Health Service Grant NS44393, and we thank Nicholas and Scott for access to the data. We thank Danielle S. Bassett for help in providing the data, use of her {\sc Matlab} code when we were debugging our code, and helpful discussions. We also thank Pawel D\l otko for useful discussions, his help with the {\sc Persistence Landscapes toolbox}, and providing us with new versions of his code during our work. We also thank Alex Arenas for helpful comments. BJS thanks the Berrow foundation for funding during her MSc degree, and BJS also gratefully acknowledges the EPSRC, MRC (grant number EP/G037280/1), and F. Hoffmann--La Roche AG for funding her doctoral studies. HAH acknowledges funding from EPSRC Fellowship EP/K041096/1.



%


\appendix


\section{Table with often-occurring brain regions in 1-dimensional loops}\label{appB}

In Table~\ref{table: Brain regions}, we indicate the brain regions that often occur in $1$-dimensional loops.

\clearpage

\begin{table*}
\caption{Loops that consist of edges that occur in loops of functional networks at least 50 times over all subjects. We list the loops that we find in the left column. (We start with one of the nodes, which we choose arbitrarily, and end with the node that is adjacent to the starting node in the loop.) 
We denote an occurrence of a loop on a specific day with the symbol $x$ in the table and present variations of the loop that we interpret as representing the same loop. We use the following abbreviations for the brain regions: \\
l: left; r: right; ant: anterior; post: posterior; AnGy: Angular gyrus; CinGy: Cingulate gyrus; COC: Central opercular cortex; FOC: Frontal operculum cortex; FMedC: Frontal medial cortex; FP: Frontal pole; HG: Heschl's gyrus; IC: Insular cortex; InfFGyPT: Inferior frontal gyrus pars triangularis; IntCalC: Intracalcrine cortex; LinGy: Lingual gyrus; OFG Occipial fusiform gyrus; OFC: Orbital frontal cortex; OP: Occipial pole; PaCinGy: Paracingulate gyrus; ParOpC: Parietal operculum cortex; PHGy: Parahippocampal gyrus; PostGy: Postcentral gyrus; PP: Planum polare; PreGy: Precentral gyrus; PT: Planum temporale; Put: Putamen; SupCalC: Supercalcrine Cortex; SuppMA: Supplemental motor area; SupMargGy: Supramarginal gyrus; SupPL: Superior parietal lobule; SupTempGy: Superior temporal gyrus; InfFGyPO: Inferior frontal gyrus pars opercularis; MTGy: Middle temporal gyrus.
}
\begin{tabular}{ |p{8cm}|p{1.5cm}|p{4cm}|p{4cm}|} 
\hline
Loop & Day $1$ & Day $2$ &Day $3$\\ \hline \hline
--lSuppMA--rSuppMA--rPreGy--lPreGy-- & x & x & x\\ \hline		 
--lOFG--lOP--rOP--rOFG-- & x & x  & x\\ \hline
--lSupTempGy ant--lPP--lHG--lPT--lSupTemGy post-- & x & x  & x\\ \hline 
--lIC--rIC--rPP--lPP-- & x & variant: --rIC--rPP--lPP--lHG--lCOC--lIC-- & variant: --rIC--rPP--rHG--lPP--lIC--  \\ \hline 
 --rIC--lIC--lPut--rPut-- & x & x & x\\ \hline
 --lIntCalC--lLinGy--lOFG--rOFG--rLinGy--rIntCalC--rSupCalC--lSupCalC-- & x &   & variant: --lIntCalC--lLinGy--rLinGy--rIntCalC--\\ \hline
--lFP--lPaCinGy--rPaCinGy--rFP-- &  & x & x \\ \hline 
--rPP--lPP--lHG--lCOC--lIC--lFOC--lInfFGyPO--lInfFGyPT--lFP--LSuppMA--lPreGy--RPreGy--rPostGy--rSupMargGyAnt--rParOpC--rPT--rSupTempGy post--rSupTempGy ant-- &  & x & variant: --rPP--rHG--rPT--rSupTempGy post--rSupTempGy ant--  \\ \hline 
 --rOFG--lOFG--lLinGy--rLinGy-- & x &   & \\ \hline
--lPaCinGy--rPaCinGy--rCinGy ant--lCinGy ant-- &  & x & \\ \hline
  --lFP--lFMedC--rFMedC--rFP-- &  & x & \\ \hline
    --lIC--lCOC--lHG--lPP-- &  & x & \\ \hline
 --lPHGy ant--lPHGy--rPHGy--rPHGy ant-- &  & x & \\ \hline
--lInfFGyPT--lInfFGyPO--lFOC--lIC--lPP--lSupTempGy ant--lSupTemGy post--lMTGy post--lMTGy ant--lFP--lOFC-- &  & x &  \\ \hline 
--rSupPL--rSupMargGy post--rSupMargGy ant--rPostGy--RPreGy-- &  & & x \\ \hline 
--rPostGy--rSupPL--lSupPL--lSupMargGy ant--rSupMargGyAnt-- &  & & x \\ \hline
--lIntCalC--lLinGy--rLinGy--rIntCalC-- &  &  & x\\ \hline
--lSupMargGy post--lAnGy--rAnGy--rSupMargGy post--rSupMargGy ant--lSupMargGy ant-- &  &  & x \\ \hline
 --lPT--lHG--lPP--lSupTempGy ant--lSupTemGy post-- &  &  & x \\ \hline \hline
Total number of loops &7 & 12& 13 \\ \hline  
\end{tabular}\label{table: Brain regions}
\end{table*}



\newpage

\section*{Supplemental Information}

\section{Topological background and definitions} \label{sec:SupportingInformation}

We give a brief introduction to the mathematical concepts behind persistent homology (PH). For our presentation, we adapt and summarize the discussion from B. Stolz's masters thesis \cite{Stolz2014}.


\subsection{Simplicial complexes} 

One can represent the underlying structures of a topological space by partitioning the space into smaller and topologically simpler pieces, which carry the same aggregate topological information as the original space when they are assembled back together. One can choose either either a small number of complicated pieces or a large number of simple pieces. From a computational point of view, the latter is preferable~\cite{Edelsbrunner2010}. 

A simple example for such a construction is the tetrahedron in Euclidian space. The tetrahedron consists of four triangular faces that are each bounded by three edges (which each connect two points). One can view the tetrahedron as a simplified version of a $2$-sphere, as it carries the same topological properties (e.g., connectedness and the enclosure of a hole) as the sphere. Similarly, one can imagine using triangles as building blocks to build more complicated constructions (e.g., ones that resemble a torus or some other manifold).

To mathematically grasp these concepts, we need a few definitions. For concreteness, we frame our discussion using the space $\R^d$ with dimension $d \in \N$.
\begin{defn}[\emph{affine combination} and \emph{affine hull}]
Let $\mathcal{U} = \{ u_0, u_1, \dots, u_k\}$ be points in $\R^d$. A point $x \in \R^d$ is an \emph{affine combination} of the points $u_i \in \mathcal{U}$, with $i \in \{0, \dots, k\}$, if there exist $\lambda_i \in \R$ such that
\begin{compactenum}[i.]
\item $x = \sum_{i = 0}^k \lambda_i u_i$\,;
\item $\sum_{i = 0}^k \lambda_i = 1$\,.
\end{compactenum}
The set of all affine combinations of $\mathcal{U}$ is called the \emph{affine hull} of $\mathcal{U}$.
\end{defn}
To ensure uniqueness of the affine combination, we introduce the following definition.
\begin{defn}[\emph{affinely independent}]
Let $\mathcal{U} = \{ u_0, u_1, \dots, u_k\}$ be points in $\R^d$. The $k+1$ points in $\mathcal{U}$ are said to be \emph{affinely independent} if the vectors $\{u_i - u_0 : \ \ i \in \left\{0, \dots ,k \}\right\}$ are linearly independent.
\end{defn}
\noindent For example, any two distinct points in $\R^2$ are affinely independent. Similarly, any three points in $\R^2$ are affinely independent as long as they do not lie on the same straight line. 

\begin{figure}[h!]
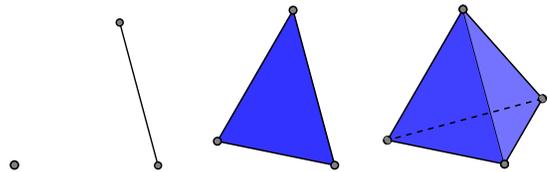

\centering\includegraphics[width=1.7cm]{Tetrahedron1.pdf} \ \ \ \
\centering\includegraphics[width=0.65cm]{Tetrahedron2.pdf} \ \ \ \ \
\centering\includegraphics[width=1.7cm]{Tetrahedron3.pdf} \ \ \ \
\centering\includegraphics[width=2.25cm]{Tetrahedron4.pdf} \ \ \ \
\caption[Examples of low-dimensional simplices.]{From left to right, we show examples of a 0-simplex, a 1-simplex, a 2-simplex, and a 3-simplex. [We adapt these examples and the figure from~[\onlinecite{Edelsbrunner2010}].]
}\label{fig:Tetrahedron2}
\end{figure}

Convex combinations and hulls are a special case of affine combinations.
\begin{defn}[\emph{convex combination} and \emph{convex hull}]
An affine combination $x = \sum_{i = 0}^k \lambda_i u_i$ is a \emph{convex combination} if $\lambda_i \geq 0$ for all $i \in \{0,\dots,k\}$. The set of all convex combinations of the points in $\mathcal{U}$ is called the \emph{convex hull} of $\mathcal{U}$.
\end{defn}
\begin{ex}
A triangle spanned by three points $u_0, u_1, u_2 \in \R^2$ is the convex hull of these points.
\end{ex}
\noindent We can now define a $k$-simplex.
\begin{defn}[\emph{$k$-simplex}]
A \emph{$k$-simplex} $\sigma = [ u_0, u_1, \dots, u_k ]$ is the convex hull of the $k+1$ affinely independent points $u_0, u_1, \dots, u_k  \in \R^d$. One calls $k$ the \emph{dimension} of the simplex.
\end{defn}
\begin{ex}\label{ex:simplex}
In Fig.~\ref{fig:Tetrahedron2}, we show examples of simplices for the first few dimensions: a point is a $0$-simplex, an edge is a $1$-simplex, a triangle is a $2$-simplex, and a tetrahedron is a $3$-simplex.
\end{ex}

\noindent The lower-dimensional simplices from example \ref{ex:simplex} are contained in the higher-dimensional simplices, because subsets of affinely independent points are also affinely independent. The lower-dimensional simplices form so-called \textit{faces} of the higher-dimensional objects.
\begin{defn}[(\emph{proper}) \emph{faces} and \emph{cofaces}]
A \emph{face} $\tau$ of a $k$-simplex $\sigma$ is the convex hull of a subset $\mathcal{V} \subseteq \mathcal{U}$. Additionally, the face is \emph{proper} if the subset relationship is a proper one. If $\tau$ is a (proper) face, then $\sigma$ is called a \emph{(proper) coface} of $\tau$.
\end{defn}
\begin{rem}
We use the notation $\tau \leq \sigma$ to denote a face of $\sigma$, and we use $\tau < \sigma$ to denote a proper face of $\sigma$.
\end{rem}

\noindent Recalling the building blocks that we described at the beginning of this Supplementary Information, we can ask whether it is only possible to build shapes using $2$-simplices (i.e., triangles) or whether one one can also combine these simplices with higher-dimensional or lower-dimensional simplices. A (permissible) shape built from a combination of simplices is called a \emph{simplicial complex}. To construct a simplicial complex, one needs to follow a set of minimal rules:
\begin{defn}[\emph{simplicial complex}]
A \emph{simplicial complex} is a finite collection of simplices $\Sigma$ such that
\begin{compactenum}[i.]
\item if $\sigma \in \Sigma$ and $\tau \leq \sigma$, then $\tau \in \Sigma$\,;
\item if $\sigma, \tilde{\sigma} \in \Sigma$, then the intersection of both simplices is either the empty set or a face of both.
\end{compactenum}
\end{defn}

\begin{figure*}[t!]
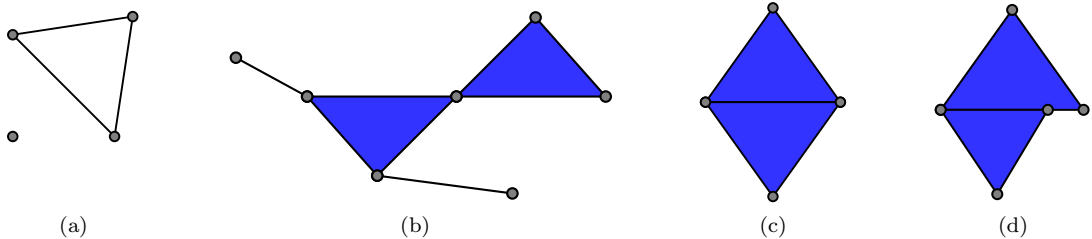

\subfloat[]{%
  \includegraphics[clip, width = 0.11\textwidth]{SimplicialComplexed0.pdf}%
}
~ 
~ 
~ 
~ 
\subfloat[]{%
  \includegraphics[clip,width = 0.3\textwidth]{SimplicialComplexed1.pdf}%
}
~ 
~ 
~ 
~ 
~ 
\subfloat[]{%
  \includegraphics[clip,width = 0.11\textwidth]{SimplicialComplexed2.pdf}%
}
~ 
~ 
~ 
~ 
~ 
\subfloat[]{%
  \includegraphics[clip,width = 0.125\textwidth]{SimplicialComplexed3.pdf}%
}
 \caption[Examples of simplicial complexes.]{Panels (a), (b), and (c) give examples of simplicial complexes. The collection of simplices in panel (d) is not a simplicial complex. We use colors to indicate $2$-simplices.}\label{fig:Simplicial2}
\end{figure*}

In Fig.~\ref{fig:Simplicial2}, we show several examples of simplicial complexes and one example that is not a simplicial complex. Example (a) illustrates that simplicial complexes are not necessarily the same as simplices. The three edges do not form a $2$-simplex, but they do form a simplicial complex that consists of $1$-simplices. In examples (b) and (c), all $1$-simplices and $2$-simplices are connected by $0$-simplices. Example (d) is a collection of simplices that violates the definition of a simplicial complex, because the intersection between the two triangles does not consist of a complete edge. Note that any combination of the three simplicial complexes (a), (b), and (c) is also a simplicial complex.

\medskip

\noindent We take the \emph{dimension} of $\Sigma$ to be the dimension of its highest-dimensional simplex. One can use simplicial complexes to represent topological spaces if there exists a homeomorphism between the simplicial complex and the topological space. Only then can one be sure that topological properties such as connectedness are preserved. 
\medskip


\subsection{Homology and Betti numbers}

\emph{Homology} is a formal way of quantitatively detecting holes in topological spaces. These holes are quantified by classifying the space that surrounds them. For example, one measures 1-dimensional holes in a torus by considering loops on its surface. One then classifies them into different types according to whether they can be deformed into each other by bending and stretching or not. In this way, one can distinguish a $2$-sphere from a torus by capturing the fact it that is possible to contract any 1-dimensional loop on the sphere to a point, whereas there are two distinct loops on the torus surface that cannot be deformed continuously into each other. These loops also cannot be contracted to a point, because they surround different holes. 

Although homology is not the only formalism that can be used for distinguishing two shapes, it currently has the fastest algorithms for computing it \cite{Edelsbrunner2010}. \emph{Homology groups}, which are topological invariants of a space, and \emph{Betti numbers} (which are derived from them) play a key role in computing homology. Homology groups detect holes in a topological space, and Betti numbers give a way to count the number of holes or distinct loops in that space.

We start constructing homology groups by looking at formal sums of simplices.
\begin{defn}[\emph{$p$-chain}]
Let $\Sigma$ be a simplicial complex, let $p$ be a given dimension, and let $G$ be an Abelian group. A \emph{$p$-chain} 
\begin{equation}
	c = \sum_{i \in I} a_i \sigma_i
\end{equation}
is a so-called ``formal sum''\cite{FormalSum} of $p$-simplices in $\Sigma$, where $a_i \in G$ are coefficients, $\sigma_i$ are $p$-simplices, and $I$ is an index set. 
\end{defn} 
\noindent In computational topology, the employed commutative group $G$ is usually $\Z/2\Z$, which has the advantage that one can regard $p$-chains as subsets of the set of all $p$-simplices in $\Sigma$ by assigning the coefficient $1$ to simplices that form part of the subset and the coefficient $0$ to those that are not in the subset. Moreover, because $\Z/2\Z$ is also a field, one can also think of $p$-chains as elements of a vector space. We use $\mathcal{C}_p = \mathcal{C}_p(\Sigma)$ to denote the set of all $p$-chains of a simplicial complex $\Sigma$.

One defines the summation of two $p$-chains, $c = \sum_{i \in I} a_i \sigma_i$ and $c' = \sum_{i \in I} b_i \sigma_i$, on $\Sigma$ in a componentwise manner: 
\begin{equation}
	c + c' = \sum_{i \in I} (a_i + b_i) \sigma_i\,.
\end{equation}
\noindent It then follows that $p$-chains form an Abelian group. When working with coefficients from $\Z/2\Z$, the sum of two $p$-chains results in summing all $p$-simplices in which the two original $p$-chains differ. The $p$-simplices that the two $p$-chains have in common are present in the sum twice, and these contributions vanish by the properties of addition on $\Z/2\Z$. 

The following definition will help relate the different $p$-chain groups of a simplicial complex.
\begin{defn}[\emph{boundary} of a $p$-simplex]
The \emph{boundary} $\partial_p\sigma$ of a $p$-simplex $\sigma = [ u_0, u_1, \dots, u_p]$ is the formal sum of its $(p-1)$-dimensional faces:
\begin{equation}
	\partial_p\sigma = \sum_{j = 0}^p [ u_0, \dots, \hat{u}_j, \dots, u_p ]\,,
\end{equation}
where $\hat{u}_j$ denotes the point that is not included when spanning the simplex.
\end{defn}
\noindent We can extend this definition to $p$-chains in a natural way by defining the boundary of a $p$-chain $c = \sum_{i \in I} a_i \sigma_i$ as $\partial c = \sum_{i \in I} a_i \partial \sigma_i$. 

We can now construct a family of boundary homomorphisms $\partial_p$ between the different groups of $p$-chains of a simplicial complex by mapping $p$-simplices to their boundaries:
\begin{align*}
	\dots \overset{\partial_{p+2}}{\longrightarrow} \mathcal{C}_{p+1} \overset{\partial_{p+1}}{\longrightarrow} \mathcal{C}_p &\overset{\partial_{p}}{\longrightarrow} \mathcal{C}_{p-1} \overset{\partial_{p-1}}{\longrightarrow} \dots \overset{\partial_{1}}{\longrightarrow} \mathcal{C}_{0} \,,\\
c &\longmapsto \partial c\,.
\end{align*}
\noindent By construction, taking the boundary of a $p$-chain satisfies the property $\partial_p(c+c') = \partial_pc + \partial_p c'$. Therefore, $\partial_p$ is a homomorphism. Such a sequence of chains and homomorphisms is called a \textit{chain complex}. One can show~\cite{Edelsbrunner2010,Croom} that the following theorem holds for boundary homomorphisms in a chain complex: 
\begin{thm}\label{thm:dd}
Let $d \in \mathcal{C}_{p+1}$. It follows that 
\begin{equation}
	\partial_p \partial_{p+1} d = 0\,.\label{eq:dd}
\end{equation}
\end{thm}

For simplicity, we often denote the boundary homomorphism by $\partial$. In other words, we omit the specification of $p$.
Two subgroups of $(\mathcal{C}_p, +)$, together with boundary homomorphisms and their property from Theorem \ref{thm:dd}, form the main ingredients in constructing the homology group of a simplicial complex. 
\begin{defn}[$p$-cycle]
A \emph{$p$-cycle} is an element of $\mathcal{Z}_p = \ker \partial_p$, where $\ker \partial_p$ denotes the kernel of $\partial_p$.
\end{defn}
\noindent We denote the set of $p$-cycles as $\mathcal{Z}_p$, and we observe that $(\mathcal{Z}_p,+)$ is a subgroup of $(\mathcal{C}_p,+)$.
\begin{defn}[\emph{$p$-boundary}]
A \emph{$p$-boundary} is an element of $\mathcal{B}_p = \text{Im } \partial_{p+1}$, where $\text{Im } \partial_{p+1}$ denotes the image of $\partial_{p+1}$.
\end{defn}
\noindent We denote the set of $p$-boundaries as $\mathcal{B}_p$, and we observe that $(\mathcal{B}_p,+)$ is a subgroup of $(\mathcal{C}_p,+)$. 

Using Theorem \ref{thm:dd}, one can now relate the cycle and boundary subgroups to each other. From Theorem \ref{thm:dd}, it follows that $\partial_p (\text{Im } \partial_{p+1}) = 0$, so $\mathcal{B}_p \subseteq \mathcal{Z}_{p}$. One can then show that $\mathcal{B}_p$ is indeed a subgroup of $\mathcal{Z}_p$.

Note that 1-dimensional loops behave differently from other edges. Edges are mapped to their end nodes by $\partial_1$, but every node in a $1$-loop occurs as the boundary of two edges and thus sums to $0$ over $\Z/2\Z$. 

We now have come very close to our goal of being able to count holes of a topological space via loops. Thus far, we have identified that the boundary subgroup $\mathcal{B}_p$ includes loops, but $\mathcal{B}_p$ may also contain the boundaries of higher-dimensional chains. To isolate the loops from the boundaries, we define the $p$th homology group of a simplex.
\begin{defn}[\emph{$p$}th \emph{homology group}]
The \emph{$p$}th \emph{homology group} $\mathcal{H}_p$ of a simplicial complex $\Sigma$ is the quotient group of the group of $p$-cycles $\mathcal{Z}_p$ modulo the group of boundaries $\mathcal{B}_p$. That is, 
\begin{equation*}
	\mathcal{H}_p =\mathcal{Z}_p / \mathcal{B}_p\,.
\end{equation*}	
\end{defn}
\noindent Two $p$-cycles in the $p$th homology group are construed as different if they differ by more than just a boundary. Otherwise, the quotient group treats them as belonging to the same homology class. Every hole of dimension $p$ in a simplicial complex is surrounded by at least one $p$-cycle in the homology group. Counting the number of classes in $\mathcal{H}_p$ thus gives an estimate of the number of $p$-dimensional loops of a simplicial complex. However, loops that surround the same hole are counted separately. A solution is to count the minimal number of elements that are needed to generate the group. 
This motivates the definition of $p$th Betti number.
\begin{defn}[\emph{$p$}th \emph{Betti number}]
The \emph{$p$}th \emph{Betti number} $\beta_p$ of a simplicial complex is 
\begin{equation*}
	\beta_p = \text{rank }{\mathcal{H}_p}\,.
\end{equation*}
\end{defn}
Recall that we are working with coefficients from $\Z/2\Z$. This turns the set of $p$-cycles into a vector space, so we can think of the homology group $\mathcal{H}_p$ as a quotient vector space. The $p$th Betti number is then given by the dimension of this vector space. One can interpret the first three Betti numbers ($\beta_0$, $\beta_1$, and $\beta_2$) to represent, respectively, the number of connected components, the number of $1$-dimensional loops, and the number of $2$-dimensional holes in a simplicial complex.


\subsection{Filtrations} 

We first define what we mean by a ``subcomplex'' of a simplicial complex $\Sigma$.
\begin{defn}[\emph{subcomplex} of a simplicial complex]
A \emph{subcomplex} of a simplicial complex is a subset of simplices that satisfy the properties of a simplicial complex.
\end{defn}
\noindent We can now build sequences of simplicial complexes that form subcomplexes of each other.
\begin{defn}[\emph{filtration}]
A \emph{filtration} of a simplicial complex $\Sigma$ is a nested sequence of subcomplexes starting with the empty complex $\emptyset$ and ending with the entire simplicial complex:
\begin{equation}
	\emptyset = \Sigma_0 \subseteq \Sigma_1 \subseteq \Sigma_2 \subseteq \dots \subseteq \Sigma_k = \Sigma\,.
\end{equation}
\end{defn}
\noindent Observe that one can define natural inclusion maps $i_j: \Sigma_j \xhookrightarrow{} \Sigma_{j+1}$ along the filtration.

In a filtration, one is interested in determining (1) when prominent features (e.g., a homology class) first appear and (2) if and when those features disappear.
\begin{defn}[\emph{birth} and \emph{death} of a homology class, \emph{persistence}]
A homology class $h \in \mathcal{H}_p(\Sigma)$ is \emph{born} at $\Sigma_m$ if $h$ is an element of $\mathcal{H}_p(\Sigma_m)$ but is not in the image of the inclusion map $i_{m-1}: \Sigma_{m-1} \xhookrightarrow{} \Sigma_{m}$. 

\medskip
\noindent A homology class $g \in \mathcal{H}_p(\Sigma)$ \emph{dies} entering $\Sigma_n$ if $g$ is an element of $\mathcal{H}_p(\Sigma_{n-1})$ but is not in the image of the inclusion map $i_{n-1}: \Sigma_{n-1} \xhookrightarrow{} \Sigma_{n}$. 

\medskip
\noindent Let $m_h$ denote the filtration step at which $h$ is born, and let $n_h$ denote the filtration step at which $h$ dies. One then defines the \emph{persistence} of a homology class $h \in \mathcal{H}_p(\Sigma)$ as
\begin{equation*}
	p_h = n_h-m_h\,.
\end{equation*}	
\end{defn}

\end{document}